\documentclass[twocolumn,showpacs,showkeys]{revtex4}
\usepackage{graphicx}
\usepackage{dcolumn}
\usepackage{bm}
\newcommand{\be}{\begin{equation}}
\newcommand{\ee}{\end{equation}}
\newcommand{\bea}{\begin{eqnarray}}
\newcommand{\eea}{\end{eqnarray}}
\def\vh{\varphi}

\begin{document}

\title{Cosmic Evolution as Inertial Motion in the Field Space of GR}

\author{ V.N. Pervushin and D.V. Proskurin
\thanks{e-mail:  proskur@thsun1.jinr.ru}}

\affiliation{Bogoliubov Laboratory for Theoretical Physics,\\
Joint Institute for Nuclear Research, 141980 Dubna, Russia}

\date{\today}

\begin{abstract}
 The identification of a cosmic scale function
 with the volume integral of a spacelike hypersurface defines
 the cosmic evolution in General Relativity
 as a collective motion along a geodesic in the field space
 of the metric components,  considered as  the coset of the affine group
 over the Lorentz one. The Friedmann equations are derived out of the homogeneous
 approximation by the Gibbs averaging
 exact equations over the relative constant spatial volume.

 A direct correspondence between  the collective cosmic motion and
 Special Relativity is established,  to  solve
 the problem of  time and energy by analogy with the solution of this problem for
  a relativistic particle  by Poincare and Einstein.
  A geometrical time interval is introduced into quantum theory of the
 relativistic collective motion by the canonical
 Levi-Civita -- type transformation in agreement with
 the correspondence principle with  quantum field theory. In this context
 the problem of quantum cosmological creation of visible matter is formulated.
 We show that latest observational data can testify to the relative
  measurement standard, and the cosmic evolution
  as an inertial motion along geodesic in the field space.
\end{abstract}



\maketitle

\newpage

\section{Introduction }

 The discovery of evolution of the Universe in the form of the Hubble law
 is considered as one of the greatest achievements of modern
 physics~\cite{f22,r28,l31,eds32}.
 The most intriguing facts are that the universe has not only the finite
 volume (to explain the Halley-de Ch\'seaux-Olbers paradox
 of the dark sky~\cite{hso,hso1,hso2}) but also the
 finite lifetime. The universe can be considered as one of ordinary physical
 objects described by  differential equations of the General Relativity (GR)
 given in a definite frame of reference~\cite{vlad} with definite  initial data and
 boundary conditions.

 The cosmic evolution in the finite space-time are conventionally described
 in the framework of the homogeneous approximation~\cite{f22,r28,l31,eds32}.
 In the present paper,   the cosmic evolution is
 considered in GR as a collective motion of  metrics in the "field space".
 The geometry of this "field space" in GR was
 obtained  by Borisov and Ogievetsky~\cite{og}
 in terms of  Cartan forms~\cite{car,dv} as a geometry of the coset
 of the affine group $A(4)$ over the Lorentz one $L$. The Cartan method
 of constructing   the nonlinear
 realization of the affine symmetry~\cite{car,dv}, in particular
 the operation of the group summation
 formulated in~\cite{vpc} -~\cite{isa}, allows us
 to extend  concepts of "collective" and "relative"
 coordinates,  inertial motions (i.e., motions with constant canonical momenta) along
 "geodesic lines" over the coset $A(4)/L$.

 These  old  concepts  extended over the field space
  reveal in GR two alternative
 measurement standards:
 the absolute standard leading to the Friedmann-Robertson-Walker (FRW)
 cosmology~\cite{f22,r28,l31,eds32}, and
 the relative standard leading to the conformal cosmology
 defined  as the FRW cosmology expressed
 in terms of conformal quantities~\cite{039,114}.  Both the measurement standards
 are discussed on an equal footing, to  compare  them with observational data.

 In the conformal cosmology we have  varying masses and constant
 temperature~\cite{N} instead of the constant masses and varying volume and
 temperature in the FRW cosmology.
  In the framework of the conformal
 cosmology both the primordial element abundance~\cite{three} and the
 latest Supernova data on the redshift -
 luminosity-distance relation~\cite{snov,riess,sn1997ff}
 are compatible with the stiff state~\cite{039}. We show here that the stiff state
 of the gravitation fields  has a simple geometric
 meaning as  the "inertial motion" of the universe along geodesic lines in
  the field coset $A(4)/L$.

 The  collective cosmic motion of the universe  (inheriting the group
 of reparametrizations of the coordinate evolution parameter in GR)
 establishes a correspondence
 between GR and Special Relativity (SR).
 The direct GR/SR correspondence  solves
  the problem of the  time and energy in GR  just as
   this problem was solved by
  Poincare and Einstein for a relativistic particle in 1904-05~\cite{poi,ein}.
 We  use modern results~\cite{grg,plb,ps1,pp,bp,bpp} on the dynamic
 description of the pure relativistic effects by the canonical
 Levi-Civita -- type transformation~\cite{lc,sh,gkp1,gkp},
 to give the mathematically rigorous introduction
 of the geometrical time interval into quantum  theory of the
 relativistic collective motion in the field space
  on the basis of the Dirac generalized mechanics~\cite{d2}.
 The Levi-Civita  canonical transformations
 convert the energy constraint into a new momentum; and
 the scale factor, into the visible (i.e., reparametrization - invariant)
  time interval measured by the watch of an observer in the universe.
 Simultaneously the Levi-Civita  transformations  convert the set of
 field variables into a new set of geometric variables with
    cosmic initial data.

 It is well  known that the quantization of cosmological models in GR
 leads to the so-called Wheeler-DeWitt equations.
 The main problem is the lost
 of time in the Wheeler-DeWitt equations.
 Just this time is restored by the Levi-Civita transformations
 that link  two evolution parameters
 and  two wave functions of the universe in the field system
 and the geometric one.
 This  Hamiltonian description of the cosmic evolution
 by two wave functions and their relation is
 the main difference of our approach from
 others (see references in~\cite{jj}).

 The GR/SR correspondence determines both the measurable energy density
 in the quantum field theory and particles as
 the holomorphic variables diagonalising this energy density.
 These variables and the mass-singularity
 of the massive vector bosons~\cite{sf,hpp}) in the Standard Models
 give theoretical basis  of solving the
 problem of quantum cosmological creation of the visible matter in
 the universe~\cite{114,yaf}

 The content of the paper is the following:
 In Section 2 we give the description of the universe
 as the collective  state of fields in the Einstein theory
 compatible with data of the observational cosmology.
  Section 3 is devoted to the reparametrization - invariant
  Hamiltonian description of the collective motion of the universe
  in both the classical theory and the quantum one.
 Section 4 is devoted to the cosmological creation of matter from vacuum.

 \section{General Relativity as a Theory of Spontaneous Breakdown
 of Affine Symmetry}

\subsection{Gravity in terms of Cartan forms of the affine group}

It is well known that the Einstein General Relativity (GR) is a gauge
field theory.
However, there is another deep analogue between gravitons in  GR
and pions in  chiral dynamics based on nonlinear
realizations of chiral symmetry $SU(2)\times SU(2)$.
In 1974 Borisov and Ogievetsky~\cite{og} showed that GR was a theory of
spontaneous breakdown
 of affine symmetry in the same way as chiral dynamics in QCD was  a theory
of spontaneous breakdown of chiral symmetry. In this theory ten
gravitons are considered as the Goldstone particles.

A theory of Goldstone particles $h$ \cite{dv,vpc,vpg,kaz,isa}
of nonlinear realizations of a semi-simple group
 G with subgroup of stability vacuum H is based on the Cartan forms
$\omega$ and $\theta$. These forms  determine Lagrangian and covariant
derivative with the
 help of a  shift $\omega$ and rotations $\theta$ of repers
 in the factor-space (i.e. coset) $K=G/H$
 defined  by the finite transformations of group
G through the equality
\be \label{cartan}
G^{-1}(h)\partial_{\mu}G(h)=i(\omega^m_{\mu}(h)X_m +\theta^n_{\mu}(h)Y_n)
\ee
with the initial data
$$
\omega^m_{\mu}(0)=0,~~~~~~~~\theta^n_{\mu}(0)=0~,
$$
where h are the group parameters identified with Goldstone particles, $Y_n(n=1,...,r)$
 are the generators of the subgroup H, and $X_m(m=1,...,k)$ are the generators of the factor space G/H. In particular,
in the exponential parametrization of group elements $G=\exp(ih^mX_m)$ the form
$\omega(h)$ describes the transition along a geodesic line in the coset
 G/H. Making the change
\be\label{sum}
G(h)~\rightarrow~G(\phi)G(h)
\ee
in (\ref{cartan}) and solving the equations for the Cartan forms with nonzero
boundary conditions \cite{vpc,vpg,kaz,isa}
\be\label{summ}
\bar
\omega^m_{\mu}(\phi,0)=\omega^m_{\mu}(\phi),~~~~~~~~\bar
\theta^n_{\mu}(\phi,0)= \theta^n_{\mu}(\phi)~,
\ee
we get the extended Cartan
forms $\bar \omega^m_{\mu}(\phi,h)$~, $\bar \theta^n_{\mu}(\phi,h)$.
These forms
describe the transition from the point $\phi$ to point $\phi(+)h$ in the
field
space, and in the particular case of the exponential parametrization, the
transition along a geodesic. The changes (\ref{sum}) and (\ref{summ})  are
called the summation in the coset.

The definition of the geodesic lines and the group summation
allow us to extend the concepts of ``inertial motion'',
the collective and relative coordinate
to the coset of the Goldstone  fields, in particular
in GR considered as a theory of nonlinear realization
of the affine group $A(4)$ in the coset $K=A(4)/L$ with respect to
the Lorentz subgroup.

The algebra of the affine group of all linear transformations
of a four-dimensional space-time $A(4)=P_4\times L(4,R)$ consists of generators
of the Lorentz group $L_{\mu\nu}$, generators of proper affine transformations
$R_{\mu\nu}$, and those of translation $P_{\mu}$.

In the theory of  nonlinear group representations
the coordinates $x_{\mu}$ and ten Goldstone fields $h_{\mu\nu}$
(i.e., gravitons) are treated as parameters of the transformations
in the factor space $A(4)/L$.
Invariants under transformations with constant parameters are constructed
with the help of Cartan forms
\bea \label{cartang}
 G^{-1}dG &=&i[\omega^P_{\mu}(d)P_{\mu }+\frac{1}{2}\omega^R_{\mu\nu}(d)R_{\mu\nu }
+\frac{1}{2}\omega^L_{\mu\nu}(d)L_{\mu\nu }]~,\nonumber\\
G&=&\exp\{iP_{\mu}x_{\mu}\}\exp\{\frac{1}{2}iR_{\mu\nu}h_{\mu\nu}\}~.
\eea
The form $\omega^R$ defines the covariant differential for the Goldstone
field $h$; and the forms $\omega^P$ and  $\omega^L$,
 the covariant diferentiation of the external field $\Psi$ transformed by
the representation of the Lorentz group with the generators $L^{\Psi}_{\mu\nu}$.
The Cartan forms are nothing but the Fock tedrads \cite{fock}
\be\label{tetra}
\omega^P_{\underline \lambda}(d)= e_{\underline \lambda \mu}dx^{\mu}~,
\ee
\be \label{tL}
\omega^L_{\underline\mu \underline \nu}(d)=
\frac{1}{2}(\omega_{\underline \mu \underline \nu}(d)-
\omega_{\underline \nu \underline \mu}(d))=
\omega_{[\underline \nu \underline \mu]}~,
\ee

\be \label{tR}
\omega^R_{\underline\mu \underline \nu}(d)=
\frac{1}{2}(\omega_{\underline \mu \underline \nu}(d)+
\omega_{\underline \nu \underline \mu}(d))=
\omega_{[\underline \nu \underline \mu]}~,
\ee
where
\be \label{tRL}
\omega_{\underline \nu \underline \mu}=de_{\underline \nu  \sigma}
(e^{-1})_{ \sigma \underline \mu}~,
\ee
and $e_{\underline \nu  \sigma}
(e^{-1})_{ \sigma \underline \mu}=\delta_{\underline \nu \underline \mu}=
{\rm diag}[1,-1,-1,-1]$.
The Cartan forms allow us to realize the Fock separation of the
Lorentz transformations
from the general coordinate ones \cite{fock}. The Fock tetrads
 $ e_{\underline \lambda \mu}$ belong to both the spaces: the Minkowski space marked by
the underlined indices $\underline \lambda$ and the Riemannian space
marked by the indices $\mu$.
The choice of the normal cordinates in the ten-dimentional space
 $h_{\underline \nu ~ \underline \mu}$ corresponds to the exponential
 parametrization \cite{og,vpg}
\be \label{normal}
e_{\underline \nu  \sigma}=(\exp h)_{\underline \nu \sigma}~.
\ee
It was shown in \cite{vpg,isa} that
the summation along geodesic lines $h(+)\phi$ corresponds to the
choice of a group transformation in the form $G=G(\phi)G(h)$ or
\be\label{tetra+}
\omega^P_{\underline \lambda}(d)= (\exp h \exp \phi)_{\underline \lambda \mu}dx^{\mu}~.
\ee
 The invariant elements of length and volume are constructed from the Cartan forms
 $\omega^P$
\bea\label{ds2}
(ds)^2=&&\omega^P_{\underline \lambda}(d)\omega^P_{\underline \lambda}(d)\equiv
\omega^P_{\underline 0}(d)\omega^P_{\underline 0}(d)\nonumber\\
& & -
\omega^P_{\underline a}(d)\omega^P_{\underline a}(d)
\equiv g_{\mu\nu}dx^{\mu}dx^{\nu}
\eea
\bea\label{dv}
dv=&&\omega^P_{\underline 0}(d)\omega^P_{\underline 1}(d)\omega^P_{\underline 2}(d)
\omega^P_{\underline 3}(d)\nonumber\\
& & =
d^4x|-e|=d^4x\sqrt{- g}~.
\eea
The four-dimensional curvature
$R=R_{\underline \mu \underline \nu ,\underline \nu\underline \mu}$,
the Riemannian tensor
\bea\label{curv}
R_{\underline \mu \underline \nu ,\underline \lambda\underline \rho   }
=&&(e^{-1})_{ \sigma \underline \lambda}
\partial_{\sigma}v_{\underline \mu \underline \nu ,\underline \rho} +
v_{\underline \mu \underline \nu ,\underline \gamma}
v_{\underline \rho \underline \gamma ,\underline \lambda}\nonumber\\
& & +
v_{\underline \mu \underline \gamma ,\underline \rho}
v_{\underline \nu \underline \gamma ,\underline \lambda}-(\underline \lambda
~\leftrightarrow~\underline \rho )~,
\eea
and the covariant differentation of the external field $\Psi$
\be\label{fock}
D_{\lambda}\Psi=D\Psi/\omega^P_{\lambda}=
 [(e^{-1})_{ \sigma \underline \lambda}\partial_{\sigma}
 +\frac{1}{2}i
v_{\underline \mu \underline \nu ,\underline \lambda}
L^{\Psi}_{\underline \mu \underline \nu}]\Psi
\ee
can be constructed from the Cartan forms \cite{og}, where
$v_{\underline \mu \underline \nu ,\underline \gamma}$ is the sum of
 \be\label{carta}
\omega_{\underline \nu \underline \mu, \underline \gamma}
=(\partial_{\lambda}e_{\underline \nu  \sigma})
(e^{-1})_{ \sigma \underline \mu}(e^{-1})_{\lambda  \underline \gamma}
\ee
 over all
permutations of the indices with the sign ($+$) for even ones; ($-$), for odd.

In terms of  these expressions
the Einstein-Hilbert action added by the Standard Model one takes the form
 \bea \label{sGR}
 S_{\rm tot}[f,e|\vh_0,M_{\rm Higgs}]= \nonumber\\
 \int\limits_{ }^{ }d^4x|-e|
 [-\frac{\vh_0^2}{6}R( e)+{\cal L}_{\rm SM}(f,e)]~,
\eea
where
\be
 \vh^2_0=M^2_{\rm Planck}\frac{3}{8\pi}~.
 \ee
One can find a direct analogue with
the action of a relativistic particle in Special Relativity (SR)
 \be
 \label{SR}
 S_{\rm SR}=-\frac{m}{2}\int\limits_{\tau_1}^{\tau_2}d\tau
[\frac{\dot X^2_{\underline \mu}}{e}+e]~,
 \ee
where the role of field variables is played by the coordinates of a particle.
 Instead of tetrads (``vier-bein'')
we have here ''ein-bein''~\cite{poi,ein}.
The analogue of general coordinate transformations
 \be \label{x}
 x_{\mu} \Rightarrow
  \tilde{x}_{\mu}=\tilde{x}_{\mu}(x_{0},x_{1},x_{2},x_{3})
 \ee
in SR is reparametrizations of the coordinate time
\be\label{tau}
\tau \rightarrow \bar \tau=\bar \tau(\tau)~.
\ee
That means that observable time is
 the invariant  interval
\be\label{s}
ds=e d\tau=\bar e d\bar \tau
\ee
  identified with the proper time
measured by the watch in the comoving frame.

 The principle of General Relativity means that coordinates of  space-time
 $x^{\mu}=(x^{0},x^{2},x^{2},x^{3})$ and tetrads $e$ in the action~(\ref{sGR})
 are not observable. Observables are the Cartan forms,
 like in electrodynamics  observables are gauge invariant tensions,
but not the gauge - variant fields.

  The Hilbert variational principle in terms of tetrads reproduces
  the classical Einstein equations
 \be\label{ee}
 \frac{\vh_0^2}{3}\sqrt{-g}\left[R_{\mu}^{\nu}(g)
 -\frac{1}{2}\delta_{\mu}^{\nu}R(g)\right]=\varepsilon_{\mu}^{\nu}
 \equiv\sqrt{-g}~T_{\mu}^{\nu}~,
 \ee
 where $T_{\mu}^{\nu}$ is  the matter energy - momentum tensor.

 The problem is to solve  equations~(\ref{ee})
 in terms of invariants of the group~(\ref{x}) in a definite {\it frame
 of reference}. The latter is defined as a set of physical instruments for
 measurement of {\it the initial data of independent
 variables}~\cite{vlad,d1,ADM,fp2,yaf1} (see Appendix A).

\subsection{Cosmic evolution as a collective motion in the coset $A(4)/L$}

 Evolution of fields including the metrics in GR is studied
 in the kinemetric frame of reference \cite{vlad} with
 the so-called Arnowitt-Deser-Misner (ADM)
  parametrization of the metric \cite{d1,ADM}
 \bea
 \label{dsevp}
  (ds)^2=&&g_{\mu\nu}dx^\mu dx^\nu= (N dx^0)^2\nonumber \\
  & &- {}^{(3)}g_{ij}
 \left(dx^i+N^idx^0\right)\left(dx^j+N^jdx^0\right)~.
 \eea

 The Hamiltonian description of GR in this frame is invariant with
 respect to reparametrization of the coordinate evolution parameter
 $x^0 \to \tilde{x}_0=\tilde{x}_0(x^0)$. This invariance
 means that the coordinate evolution parameter $x^0$ cannot be  measured,
 and we should point out a measured evolution parameter among the
 set of the field variables to solve the problem of energy and time  in
 the kinemetric frame of reference \cite{pp,bpp,tmf}.
 This measured evolution parameter is well known in cosmology
 as the cosmic scale factor.

 In contrast to  the conventional homogeneous
 approximation \cite{f22}-\cite{eds32}, we define
 the cosmic scale factor $a(x^0)$
 as the functional of metrics in the form of an invariant spatial volume
 \bea\label{volum}
 \frac{1}{V_0}\int\limits_{V_0}d^3x\sqrt{{}^{(3)}g(x^0,x^i)}
 \equiv\frac{1}{V_0}\int\limits_{V_0}d^3x
 |{}^{(3)}e|= a^3(x^0),
 \eea
 where $V_0$ is the present-day value of the finite volume.

 In contrast to  the conventional Hamiltonian description of
 GR \cite{d1,ADM,fp2},
 we extract the cosmic scale factor  as a collective coordinate
  in the space of metrics $g_{\mu\nu}$ considered as the coset
 A(4)/L~\cite{og}. We introduce the collective
 variable using the geometry of geodesic lines in  the coset $A(4)/L$.
 The operation of adding along a geodesic line in terms of
 the normal coordinates in the field space
 $g_{\mu\nu}(h)=[\exp(2h)]_{\mu\nu}$
 is defined by eq. (\ref{tetra+})~\cite{vpg,isa}:
 \bea \label{vp}
  g_{\mu\nu}(h_{\rm coll.}(+)h_{\rm rel.})=&& \nonumber\\
  \{\exp(h_{\rm coll.})\exp(2h_{\rm rel.})\exp(h_{\rm coll.})\}_{\mu\nu}~.
 \eea
 In this case, a collective motion of the volume $a(x^0)$ is separated
 from  the relative metric $\bar g_{\mu\nu}(x^0,x^i)$ by
 the multiplication
 \be\label{glob}
 g_{\mu\nu}(x^0,x^i)=\bar g_{\mu\nu}(x^0,x^i)a(x^0)^2
   \ee
 corresponding to
 $$
 N=\bar N a,~
 g_{ij}=\bar g_{ij}a^2~.
 $$
The normal
 coordinate in the field space along a geodesic line
 is the Misner exponential parametrization of the scale factor~\cite{M}
 \be\label{mn}
 a(x^0)=\exp X_0(x^0)~.
  \ee
 Constant values of the canonical momentum of the Misner
  variable $X_0$ correspond to an {\it inertial motion} in the
field space along a geodesic line.

  Transformation (\ref{glob}) is a particular case of the Lichnerowicz
 conformal transformations~\cite{L}
 of all field variables $\{{}^{(n)}f\}$
 \be\label{lic}
 {}^{(n)}f(x^0,x^i)={}^{(n)}\bar f(x^0,x^i) a(x^0)^n
 \ee
 with a conformal weight $n$, including
 the metric as a tensor field with the conformal weight
 $n=2$ ~\cite{Y}.
 We suggest that  each field contributes to the cosmic  evolution
 of the universe in line with the Lichnerowicz
 conformal transformations (\ref{lic}).  The auxiliary
 variable can be  removed by the  constraint
 of the constant spatial volume in
 the relative field space $\bar g_{\mu\nu}$ that follows from eqs. (\ref{volum}),
 (\ref{glob})
 \be\label{volume}
 \int\limits_{V_0}d^3x|{}^{(3)}\bar e(x^0,x^i)|\equiv V[\bar e]=V_0~.
 \ee
 To identify  this {\it collective variable}
 with the homogeneous cosmic scale factor in observational cosmology,
 we  should verify that the exact equations of $a(x^0)$ averaged over
 the  invariant three-dimensional
  volume in our theory  coincide with the equations of
  homogeneous cosmic scale factor in
 the standard cosmology where the concept of the cosmic evolution of
 the universe is formulated.

 \subsection{Exact equations of the cosmic evolution }

 To find the Einstein action with the collective motion
 and corresponding Einstein equations, we use the
 well-known formula of conformal transformations~(\ref{glob})
 of a four-dimensional curvature
 \be\label{ctr}
 \sqrt{-g}\frac{\vh_0^2}{6}R(g)=\sqrt{-\bar g}\frac{\vh^2}{6}R(\bar g)
 -\vh\partial_{\mu}
 \left[\sqrt{-\bar g}\bar g^{\mu\nu}{\partial_{\nu}\vh} \right],
 \ee
 where $\vh(x^0)$ is the dynamic Planck mass defined as the product
 of the Planck mass and the cosmic scale factor
 \be\label{vh}
 \vh(x^0)=a(x^0)\vh_0~~~~~~~~~~~~
 \left(\vh_0=M_{\rm Planck}\sqrt{\frac{3}{8\pi}}\right).
 \ee
  This formula leads to the Einstein action
 \bea\label{GRc}
 S_{GR}[e|\vh_0]=&&
 S_{GR}[\bar e|\vh]\nonumber\\
&&+
\int\limits_{x^0_1 }^{x^0_2 }dx^0
 \int\limits_{V_0}d^3x\vh
 \frac{d}{dx^0}
 \left(|{}^{(3)}\bar e|\frac{d \vh }{\bar N dx^0}\right)\nonumber\\
&&+\int\limits_{x^0_1 }^{x^0_2 }dx^0\Lambda(x^0)\left[V[\bar e]-V_0
\right]~,
 \eea
 where  the Lagrangian factor $\Lambda(x^0)$ provides the conservation
 of the volume~(\ref{volume}) of  local excitations;
 \bea \label{Econf}
 S_{\rm GR}[\bar e|\vh]=
 -\int d^4x\sqrt{-\bar g}\frac{\vh^2}{6}R(\bar g) \nonumber\\
 = \int\limits_{x^0_1 }^{x^0_2 }dx^0\int\limits_{V_0}d^3x
 \left[ \verb"K"(\bar e|\vh)-\verb"P"(\bar e|\vh)
 +
 \verb"S"(\bar e|\vh)\right]
\eea
 is the standard ADM action in GR with the relative metric $\bar g$ and the
 running Planck mass where
 \be \label{Eck}
 \verb"K"(\bar e|\vh)=\frac{\vh^2|{}^{(3)}\bar e|}{24 N}
[\bar \pi_{\underline a \underline b}\bar \pi_{\underline a\underline b}-
 \bar \pi_{\underline b\underline b}\bar \pi_{\underline a\underline a}]
 \ee
 is the kinetic term with the external form
$$
\pi_{\underline a \underline b }=
\frac{1}{2}
[(D_0e)_{\underline a i}(e^{-1})_{i\underline b}+(
\underline a~\leftrightarrow~\underline b)])~,
$$
 \be \label{extovp}
(D_0e)_{\underline a i}=\left(\partial_0-N^l\partial_l\right)
e_{\underline a i} -e_{\underline a l}\partial_iN^l~,
\ee

 \be \label{Ecp}
 \verb"P"(\bar e|\vh)=\frac{\vh^2
 \bar N |{}^{(3)}\bar e|}{6}~~{}^{(3)}R({}^{(3)}\bar g)
 \ee
 is the potential term, and
 \bea \label{EcS}
 \verb"S"(\bar e|\vh)=&&\frac{\vh^2 }{6}\left(\partial_0
 - \partial_kN^k\right)
 \left(|{}^{(3)}\bar e|
 \frac{\bar \pi_{\underline a\underline a}}{\bar N}\right) \nonumber\\
&&
-\frac{\vh^2 }{3}
 \partial_i\left(|{}^{(3)}\bar e|{}^{(3)}
 \bar g^{ij}\partial_j \bar N\right)
 \eea
 are   the standard
 ADM "surface terms"  contributing to the equations of motion
 due to the time dependence of the dynamic Planck mass $\vh$.
 This contribution describes the interference between
 the local relative excitations and the collective one.

 The action of collective motion allows us to define
 the global lapse function
  \be\label{N2}
 \frac{1}{\bar N_0(x^0)}= \frac{1}{V_0}\int\limits_{V_0}d^3x
 \frac{ |{}^{(3)}\bar e|}{ \bar N}.
 \ee
 and the gauge - invariant {\it world geometric time}
 \be\label{ct}
 d\eta=\bar N_0(x^0)dx^0= \tilde {\bar N}_0(\tilde x^0)d\tilde x^0~.
 \ee
 In terms of {\it world geometric time}
the variation of the total action~(\ref{sGR}) with respect to
 the lapse function
 and determinant of spatial metric leads to the equations
 \bea\label{ladm00g}
 \bar N\frac{\delta S_{\rm tot}}{\delta \bar N }
 =0&|\!\!\models\!\!\!\!\!\!
 \Longrightarrow& \frac{\vh'^2}{{\cal N}}=
 {\cal N}\bar T_0^0,
 \\
 \label{ladmkkg}
 \bar g^{ij}\frac{\delta S_{\rm tot}}{\delta \bar g^{ij} }=
 0&|\!\!\models\!\!\!\!\!\!\Longrightarrow&
\frac{ 2(\vh^2)''-{3}\vh'^2}{\cal N} +3\Lambda=
{\cal N}\bar T^k_k,\\
\frac{\delta S_{\rm tot}}{\delta \bar N^k }=0&|\!\!\models\!\!\!\!\!\!
 \Longrightarrow& \bar T_k^0=0,
 \\
 \bar g^{ki}\frac{\delta S_{\rm tot}}{\delta \bar g^{ij} }=0&|\!\!\models\!\!\!\!\!\!
 \Longrightarrow& \bar T_k^i=0~~(i\neq k),
  \eea
where  $f'=df/d\eta$, ${\cal N}=\bar N/\bar N_0$, and $\bar T_{\mu}^{\nu}=
T_{\mu}^{\nu}-\vh^2/3(R_{\mu}^{\nu}-1/2\delta_{\mu}^{\nu}R)$ are the total
components of the local energy-momentum tensor
 \bea\label{1ladm00g}
  |{}^{(3)}\bar e|{\cal N}\bar T_0^0
 =\verb"K"(\bar e|\vh)+\verb"P"(\bar e|\vh)+\varepsilon^0_0\equiv
 \varepsilon^0_{0({\rm tot})}~,
 \eea

 \bea \label{1ladmkkg}
 3\verb"K"(\bar e|\vh)
 -\verb"P"(\bar e|\vh)+2\verb"S"(\bar e|\vh)+ \varepsilon^k_k\equiv
 \varepsilon^k_{k({\rm tot})}
 \eea
($\bar T_{\mu}^{\nu}=0$ is equal  to zero, if the cosmic evolution is absent $\vh(x^0)\equiv\vh_0$).
 These equations contain the
 collective motion of the cosmic evolution that can be extracted by
 integration of these equations over the spatial volume.
 As a result we get
 \bea\label{ladm00h}
 \frac{1}{V_0}\int\limits_{V_0}d^3x\bar N
 \frac{\delta S_{\rm tot}}{\delta \bar N }
 =0&|\!\!\models\!\!\!\!\!\!\Longrightarrow&
 \vh'^2=\rho_{\rm tot},
 \eea
  \bea\label{ladmkkh}
 \frac{1}{V_0} \int\limits_{V_0} d^3x\bar g^{ij}\frac{\delta S_{\rm tot}
 }{\delta \bar g^{ij} }=0&|\!\!\models\!\!\!\!\!\!\Longrightarrow&
 (\vh^2)''-{3}\vh'^2+3\Lambda = \nonumber\\
  -3p_{\rm tot}~;
 \eea
 here we introduce the Gibbs averaging
 $$
 \rho_{\rm tot}=\frac{1}{V_0}\int d^3x\varepsilon^0_{0({\rm tot})}~,
 $$
 $$
 3p_{\rm tot}=\frac{1}{V_0}\int d^3x\varepsilon^k_{k({\rm tot})}.
 $$
 These equations are accompanied by the equations of collective
 variables
 \bea
 \label{ladmkki}
 \vh\frac{\delta S_{\rm tot}}{\delta \vh }
 =0&|\!\!\models\!\!\!\!\!\!\Longrightarrow&
 2\vh\vh''
 =\rho_{\rm tot}-3p_{\rm tot},\\
  \label{ladmkkj}
 \frac{\delta S_{\rm tot}}{\delta \Lambda }=
 0&|\!\!\models\!\!\!\!\!\!\Longrightarrow&
 V{[\bar g]}-V_{0}=0.
 \eea
 The combination of eqs. (\ref{ladm00h}), (\ref{ladmkkh}), and (\ref{ladmkki})
 leads to $\Lambda=0$.

 In this case, the exact equations~(\ref{ladm00h}) and~(\ref{ladmkkh}) in
 {\it relative field space} for the collective variable
 completely coincide with the  conformal version of the equations
 of the Friedmann-Robertson-Walker (FRW) cosmology in
  the homogeneous approximation \cite{f22}-\cite{eds32}
 \be \label{SCe}
\begin{array}{|c|} \hline \\
 \vh_0^2 a'^2=\rho_{\rm tot}~;
 ~~~\vh_0^2 \left[3a'^2-(a^2)''\right]=3p_{\rm tot}~,
 \\ \\ \hline
\end{array}
\ee
 where $\rho_{\rm tot}$ and $p_{\rm tot}$
 are the  total density and the
 total pressure.
Transition to physical values (time $t$,
distance $l$,
density $\rho_{\rm F}$) of the FRW cosmology is carried out with
the help of conformal transformations
\bea\label{preobr}
t&=&\int_0^\eta d\bar\eta a(\bar\eta),\\
l&=&a(\eta)r_{\rm},~~r_{\rm}=\sqrt{x_1^2+x_2^2+x_3^2}\\
\rho_{\rm F}(a)&=&\frac{\rho_{\rm tot}(a)}{a^4}~.
\eea
In the terms of the FRW cosmology the equation of evolution~(\ref{SCe}) takes
the conventional form
\be\label{preobrSK}
\begin{array}{|c|} \hline \\
 \vh_0^2 \left(\frac{da}{dt}\right)^2=\rho_{\rm F}(a)~;
 ~~~\vh_0^2 \left[\left(\frac{da}{dt}\right)^2+a\frac{d^2a}{dt^2}\right]
 =-3p_{\rm F}.
 \\ \\ \hline
\end{array}
\ee

 In addition, by  substituting eqs.~(\ref{ladm00h})
 and~(\ref{ladmkkh})
 into~(\ref{ladm00g}) and~(\ref{ladmkkg}),
 we obtain the equations of the local excitations
 \be \label{lee}
 \frac{|{}^{(3)}\bar e|}{{\cal N}V_0}\int\limits_{V_0} d^3x
 \varepsilon^0_{0\mbox{(\rm tot)}}=
 \varepsilon^0_{0\mbox{(\rm tot)}}~,
 \ee
 $$\frac{|{}^{(3)}\bar e|}{{\cal N}V_0}\int\limits_{V_0} d^3x
 \varepsilon^k_{k\mbox{(\rm tot)}}=
 \varepsilon^k_{k\mbox{(\rm tot)}}~,
 $$
 where $\varepsilon^0_{0\mbox{(\rm tot)}},~\varepsilon^k_{k\mbox{(\rm tot)}}$
 are given by eqs.~(\ref{1ladm00g}) and~(\ref{1ladmkkg}).\par
 These local equations are compatible with the
 cosmological equations~(\ref{ladm00h}),~(\ref{ladmkkh}),
 and in the infinite volume limit the local equations~(\ref{lee})
 coincide with the ordinary Einstein equations in
 the Riemannian space-time.
It was shown that the cosmic evolution changed the Newton law and the
black-hole solution in the Early Universe~\cite{tmf,114}.

\subsection {Measurement standards}

 It is worth reminding that the concept of measurable quantities
 in the field theory is no less important than
  the equations of the theory. J.C. Maxwell wrote:
 ''The most important aspect
 of any phenomenon from  mathematical
 point of view  is that of a measurable quantity.
  I shall therefore consider electrical phenomena
  chiefly with a  view to their measurement,
 describing the methods of measurement, and
 defining the  standards on
  which they depend''\cite{we}.

Suppose that nature selects itself both the theory and
standards of measurement, and the aim of observation is
to reveal not only initial data, but also these measurement standards.
In particular, one of the
central concepts of the modern cosmology is the concept of the scale defined
as a functional of spatial volume in GR~\cite{tmf}.
If expanding volume of the universe means
the expansion  of "all its lengths",
we should specify whether the measurement standard of length
expands.  Here there are two possibilities: the first, the absolute
  measurement standard does not expand;
  and the second, the relative measurement standard
 expands together with the universe.

Until the present time the first possibility was  mainly considered
in cosmology. The second possibility means that
we have no absolute instruments to measure absolute values
 in the universe. We can measure only a ratio of values which does not
 depend on the spatial scale factor. The relative measurement standard
 transforms the spatial scale of the intervals of lengths into
 the scale of masses which permanently grow.

 As we have shown in the previous section the universe evolution as a collective motion of
 the spatial volume in the field "space"
$$
(\vh,~\bar g_{\mu \nu}, \bar f...)\equiv (\vh,~\bar F)
$$
 reproduces the equations of a conformal version of the FRW cosmology,
 if the homogeneous approximation
 is changed by  averaging the local density and pressure.
 To obtain the equation of the standard cosmology, it
 is sufficient to make the reverse conformal transformations~(\ref{lic}) of
 the {\it relative} quantities ${{}^{(n)}\bar F}$
 into the quantities of {\it absolute field space}
  ${}^{(n)}F={}^{(n)}{\bar F}a^n$.
 The FRW cosmology supposes
 that our instruments measure absolute fields ${}^{(n)}F$ and
 a "absolute" interval of the Riemannian
 space
 \be \label{dse1}
  (ds)^2=g_{\alpha\beta}dx^\alpha dx^\beta~.
 \ee
  The "relative" point of view supposes
 that our instruments measure relative fields ${}^{(n)}\bar F$ and
 a "relative" interval of the Riemannian
 space
 \be \label{dse10}
  (d\bar s)^2= \bar g_{\alpha\beta}dx^\alpha dx^\beta\equiv
\frac{ (ds)^2}{a^2}~.
 \ee

 \subsection{The GR/SR  Hamiltonian correspondence}

 The conventional Hamiltonian formulation
in terms of Dirac theory of constrained systems~\cite{d2}
is given by the action

\bea\label{1hallu}
S_{\rm GR}=~~~~~~~~~~~~~~~~~~~~~~~~~~~~~~~~~~~~~~~~~~~~~~~~~~~~~~~~~ \nonumber\\
 \int dx^0\left\{
\left[ \int d^3x \sum\limits_{i,\underline{a}}
P_{i\underline{a}}\partial_0\bar e^i_{\underline{a}}\right]-
H_{\rm tot}[\bar e |\vh]
\right\}~,
\eea
where
 \bea\label{1Hvp}
H_{\rm tot}[e|\vh_0]=~~~~~~~~~~~~~~~~~~~~~~~~~~~~~~~~~~~~~~~~~~~~~~~~ \nonumber\\
\int d^3x\left[
N{\cal H} -N^k {\cal P}_k + C_0P_{g}+C_{\underline b}f_{\underline b}( e)-
\verb"S"\right]
\eea
is the  Hamiltonian,
  $ N, N^k$,
 $C,~C^k$ are the Lagrangian multipliers for
 the first class constraints  for densities of energy ${\cal H}=0$ and
  momenta $ {\cal P}_k=0$,
 and second class ones that are the Dirac conditions of
   transverseness
 $f_{\underline b}( e)=0$ and
 the minimum embedding of three-dimensional hypersurface into
 the four-dimensional Riemannian space-time
 with the zero momentum of spatial metric determinant
 $P_g=0$~\cite{d1}.
 The latter means that the second (external) form
 $\pi_{\underline{a},\underline{a}}$
  is equal to zero that contradicts to
 the cosmic evolution with the nonzero Hubble parameter proportional to
 $\pi_{\underline{a},\underline{a}}$ (see Appendix A, eq.~(\ref{hall1})).

On the other hand, the conventional description of the cosmic evolution keeps
only the homogeneous part of the second form
neglecting all local excitations $H_{\rm tot}=0$.

To include the cosmic evolution into field theory of the local
excitations, we have define this evolution~\cite{pp,plb} as the collective
variable $\vh(x^0)=a(x^0)\vh_0$ by eq. (\ref{volum}).

 The Einstein theory after  the separation of
 the collective motion takes the form
 \be \label{act}
 S_{\rm GR}[e|\vh_0] = S_{\rm GR}[\bar e|\vh] + S_{\rm interference} +
 S_{\rm collective}~,
 \ee
 where the first term coincides with the initial Einstein action
in terms of relative fields and {\it dynamic evolution parameter} $\vh$
instead of the Planck mass $\vh_0$; the second term
 \be \label{interf}
S_{\rm interference} = - \frac{1}{6}\int\limits_{x^0_1 }^{x^0_2 }dx^0
{\partial_0 (\vh^2) } \int
 \limits_{V_0}d^3x \frac{\bar \pi_{\underline a\underline a}}{\bar N}
 \ee
goes from the first one in eq.~(\ref{EcS})
in the relative metric. This term
describes an interaction of the collective and relative variables.
The third term
 \be \label{universe}
S_{\rm collective} =  - V_0\int\limits_{x^0_1 }^{x^0_2 }dx^0
\frac{(\partial_0 \vh)^2 }{\bar N_0}
 \ee
is the collective motion of the universe.

 The interference of the collective and relative variables disappears,
 if we impose the Dirac condition~\cite{d1}
 of the minimal embedding of the three-dimensional hypersurface
 into the four-dimensional space-time in the relative space
 \be\label{mem}
 \bar \pi_{\underline a\underline a}=  0~.
 \ee
 The minimal embedding removes not only the interference
 of the collective motion with local excitations, but also
 all local excitations with the negative norm~\cite{fp2}.

 In the case of the minimal embedding the Hamiltonian form of
 the  Einstein action~(\ref{1hallu})
 with the  collective motion takes the form

\begin{widetext}
\bea
 \label{hallu}
S_{\rm GR}[e|\vh_0]=\int dx^0
\left[ \int d^3x \sum\limits_{i,\underline{a}}
P_{i\underline{a}}\partial_0\bar e^i_{\underline{a}} -
H_{\rm tot}[\bar e |\vh]
-P_{\vh}\partial_0\vh +\bar N_0\frac{P_{\vh}^2}{4V_0}+
\Lambda (\bar V[\bar e]-V_0)\right]
~;
\eea
\end{widetext}
here the relative Hamiltonian $H_{\rm tot}[\bar e |\vh]$
 is the conventional one~(\ref{1Hvp})
 where  all fields are changed by the relative one
  $\bar e^i_{\underline{a}},\bar N$
and the Planck mass $\vh_0$ is changed by the running
Planck mass as {\it dynamic evolution
parameter} $\vh$, $\bar N_0[\bar e,\bar N]$
 and $\bar V[\bar e]$ are considered as
   functionals given by eqs.~(\ref{volume}) and~(\ref{N2})
 \be\frac{1}{\bar N_0}=\frac{1}{V_0}\int\limits_{V_0}d^3x
 \frac{ |{}^{(3)}\bar e|}{\bar N},
 \qquad \bar V[\bar e]=\int\limits_{V_0}d^3x|{}^{(3)}\bar e|~.
 \ee

The GR with the collective motion is the direct field generalization of
SR with two time-like variables (the geometric interval $d\eta=N_0dx^0$ and
{\it dynamic evolution parameter} $\vh$) and two wave functions.
We have one to one correspondence between SR and GR~\cite{ps1,pp},
i.e., their proper times
\be\label{corrt}
         ds=ed\tau~~~~~~~~\Longleftrightarrow ~~~~~~~~~
          d\eta=\bar N_0dx^0~,
\ee
their world spaces
\be\label{corrw}
         X_0,~X_i~~~~~~~    \Longleftrightarrow
         ~~~~~~~~~~  \vh,~\bar e^i_{\underline{a}}~,
\ee
their energies
\be\label{corre}
 P_0=\pm\sqrt{P_i^2+m^2}~ \Longleftrightarrow
 ~ P_{\vh}=
\pm2\sqrt{V_0H_{\rm tot}[\bar e |\vh]},
\ee
and their two-time   relations in the differential form
\be\label{corrdt}
 \frac{dX_0}{ds} =\pm\frac{\sqrt{P_i^2+m^2}}{m}~
\Longleftrightarrow ~
\frac{d\vh}{d\eta} =
\pm \sqrt{\rho_{\rm tot}(\vh)}
\ee
and in the integral forms
\begin{widetext}
\bea\label{corrit}
  s(X_0)=\pm\frac{m}{\sqrt{P_i^2+m^2}}X_0  ~\Longleftrightarrow
\eta(\vh_0,\vh_I) =
\pm \int\limits_{\vh_I}^{\vh_0}\frac{d\vh}{\sqrt{\rho_{\rm tot}(\vh)}}~.
\eea
\end{widetext}

Recall that in SR eq. (\ref{corrit}) can be treated as the Lorentz
 transformation of the rest frame with time $X_0$ into the comoving one
 with the proper time $s$. Similarly, in GR  the relation
 $\eta(\vh)$ defined by eq. (\ref{corrit}) is treated
  as a canonical transformation \cite{pp}. This GR/SR correspondence
  (\ref{corrt})-(\ref{corrit}) allows us to solve the problem of time
  and energy  in GR  like Poincare and Einstein \cite{poi,ein}
   had solved it in SR. They identified the time with one of variables
   in the world space. The similar String/SR correspondence was
   considered in papers \cite{bp,bpp,tmf}.

\subsection{Cosmic evolution as an inertial motion in the coset $A(4)/L$}

The cosmic collective motion as the dynamics of the scale factor $a$
can be separated in any theory, in particular,
in the unified theory considered as the sum of
 GR and the Standard Model
 \bea\label{tot}
 S_{\rm tot}[\{F\}|\{M\}]=~~~~~~~~~~~~~~~~~~~~~~~~~~~~~~~~~~~~~~~~~ \nonumber\\
 S_{\rm GR}[e|M_{\rm Planck}]+
 S_{\rm SM}[e,\{f\}|M_{\rm Higgs}]
 \eea
 with a set of fields $\{F\}=e,\{f\}$ and a set of massive parameters
 $\{M\}$  including the Higgs mass. This separation
  is fulfilled by the Lichnerowicz transformations of fields
  with the conformal weight $n$ ${}^{(n)}F={}^{(n)}  \bar F a^n$~(\ref{lic}).
  As the result, the action~(\ref{tot}) takes the form
 \bea\label{1slic}
 S_{\rm tot}[\{F\}|\{M\}]=~~~~~~~~~~~~~~~~~~~~~~~~~~~~~~~~~~~~~~ \nonumber\\
 S_{\rm tot}[ \{\bar F\}|\{M a\}]+
 S_{\rm collective}[a,\bar N_0]~,
 \eea
 where
\be\label{2slic}
S_{\rm collective}[a,\bar N_0]= -\gamma\int\limits_{x^0(I)}^{x^0(0)}dx^0
\left[\frac{(\partial_0a(x^0))^2}{\bar N_0}\right]~,
\ee
 where $\gamma=V_0\vh_0^2$.
To make the analogue with a relativistic particle
more transparent, we can pass to
the normal coordinates in the field space along a geodesic line
that corresponds to the choice of the Misner  parametrization
of the evolution
 parameter $X_0=\log a$ in the field space
  (\ref{mn}) and the Misner lapse function
 \be\label{uncol0}
 a=e^{X_0};~~~~~~~~~~~~~~
 ~~e_{0} = \bar N_{0}e^{-2 X_0 }~.
 \ee
Then, instead of  the action~(\ref{2slic}) we get
 \be\label{uncol10}
 S_{\rm collective}[X_0,N_0]=- \gamma
 \int\limits_{x^{0}_{1}}^{x^{0}_{2}} dx^{0}
 \left[ \frac{(\partial_{0}X_{0})^2}{e_0} \right]~.
 \ee
  We call the collective motion
 along the geodesic line  {\it inertial},
 if the canonical momentum of the field evolution parameter $X_0$
 \be\label{canun}
 P_0=-2\gamma\frac{{(\partial_{0}X_{0})}}{e_0}
 \ee
 is a constant
 \be\label{canun1}
 \frac{d P_0}{d x^0}=-2\gamma\frac{d}{d x^0}
 \left[\frac{{(\partial_{0}X_{0})}}{e_0}\right]=0
 ~\Longrightarrow~\frac{{dX_{0}}}{e_0dx^0}=H_0~.
 \ee
The solution of this equation
 is expressed in terms of the invariant "Misner time"
 $d\Omega=e_{0}dx^0$
 \be\label{gsp2}
  X_0(\Omega)=H_0\int\limits^{x^0} e_0(\bar x^0) d\bar x^0=H_0\Omega(\eta)~.
 \ee
 The dependence $\Omega(\eta)$ on the conformal time
 $d\eta=\bar N_{0}dx^0$~(\ref{ct}) follows from  eq.~(\ref{uncol0})
 \be \label{gsp}
 e_{0}dx^0=d\Omega = e^{-2 X_0(\Omega) }d\eta~.
 \ee
 The solution of this equation takes the form
 \be\label{gsX}
 X_0(\Omega)= H_0\Omega(\eta)\equiv
 \frac{1}{2}\log[1+2H_0(\eta-\eta_0)]~.
 \ee
  In this case, the scale factor $a=\exp (X_0)$ is proportional
  to the square root of the
  conformal time~(\ref{ct}) $d\eta=\bar N_0dx^0$
\be\label{evol}
 a^2(\eta) =  [1 + 2H_0(\eta-\eta_0)]~,
 \ee
where $\eta_0$ is the present-day value of time $a(\eta_0)=1$.
This result follows directly from the equations of motion~(\ref{SCe})
if the pressure is equal to the density
 \be \label{pr}
 p_{\rm tot}(a) = \rho_{\rm tot}(a)=\frac{\vh_0^2H_0^2}{a^2}
 \ee
and these equations reduce to
\be\label{1evol}
(a^2)'' = 0 ~.
 \ee
 Thus, the {\it inertial} cosmic
 motion corresponds to the so-called stiff state~(\ref{pr}).
 As it is known from the observational cosmology,
 the standard  matter  $S_{\rm tot}[ \{\bar F\}|\{M e^{X_0}\}]$
 gives a small contribution to the cosmic evolution.

 Therefore, we propose that the equation of the stiff state
 is described by an additional  action $S_I[e_0]$ so that the complete
 action takes the form
 \bea\label{3slic}
 S_{\rm tot}[\{F\}|\{M\}]+S_I[e_0]=&&
 S_{\rm tot}[ \{\bar F\}|\{M_Ie^{X_0}\}]\nonumber\\
&&+
 S_{\rm collective}[X_0,e_0]\nonumber\\
&&+
S_I[e_0]~.
 \eea
 If we neglect the first term in~(\ref{3slic}),
 the additional  action $S_I[e_0]$ leads to
 the inertial motion of the universe along geodesic
 with the density~(\ref{pr})
\bea\label{1unco2}
 S_{\rm universe}&=&S_{\rm collective}[X_0,N_0]+S_I[e_0]
 =\nonumber\\
&-& \gamma
 \int\limits_{x^{0}_{1}}^{x^{0}_{2}} dx^{0}
 \left[ \frac{(\partial_{0}X_{0})^2}{e_0} +e_0H_0^2\right]~.
 \eea
This action describes the relativistic universe
in which the problem of energy is solved like in Special
Relativity~(\ref{SR}).

Thus, there are two differences of the cosmic motion in the coset
$A(4)/L$ from the standard cosmology in the FRW metrics.
These are the reparametrization invariance and the relative measurement standard
in the coset which leads to the {\it conformal cosmology} with a constant volume of the
flat space
$$
d\bar s^2=d\eta^2-dx^idx^i,~~~~~~~~~~~~r^2=x^ix^i\equiv (x^1)^2+(x^2)^2+(x^3)^2
$$
and varying masses $\bar M(\eta)=M a(\eta)$ defined by eq.~(\ref{evol}).
Therefore,  the spectrum of atoms is described by the Schr\"odinger equation
\be\label{Sch1}
\left[\frac{\hat p^2}{2m_0
 a(\eta)}-\left(\frac{\alpha}{r}+E(\eta) \right) \right]\Psi_A=0~.
 \ee
 It is
 easy to check that the exact solution of this equation is expressed throw the
 solution $E_0$ of a similar Schr\"odinger equation with constant masses $m_0$
 at $a(\eta_0)=1$
 \be\label{Sch2}
 E(\eta)=a(\eta)E_0\equiv\frac{E_0}{z(d)+1}~,~~~~~~~~~~~~~ E_0=-\frac{
 m_0\alpha^2}{n^2}~,
 \ee
  where $z(d)$
 is a redshift of the spectral lines of
 atoms at the coordinate distance $d/c=\eta_0-\eta $, and $\eta_0$ is the
 present-day value of the geometric (conformal) time.

\subsection{Conformal cosmology and SN data}

 This type of
 conformal cosmology was  developed by Hoyle and Narlikar~\cite{N}.
 A red photon emitted by an atom at a star two billion years
 (in terms of $\eta$)
 remembers the size of this atom, and after two billion years this
 photon is compared with a  photon of
 the standard atom at the Earth that became blue due to
 the evolution of all masses.
 The redshift-coordinate distance relation
 is defined by the formula of the standard cosmology~(\ref{Sch2})
 \be \label{reds}
 z(d)={a(\eta_0)\over a(\eta_0-d/c)}-1~,~~~~~~a(\eta_0)=1~%
 \ee
 (where $d$ is the coordinate distance to an object)
 because the description
 of the conformal - invariant photons does not depend on the
 standard of measurements.

In the case of the inertial motion~(\ref{evol})
 this redshift - distance relation
takes the form
\be\label{cldr1}
z(d)={1\over \left(1+2H_0d/c\right)^{1/2}}-1~.%
\ee
It results in the following simple relation
\be\label{cldr2}
d(z)=\frac{c}{2H_0}\left[1-{1\over (1+z)^2}\right]~.
\ee
  The redshift - luminosity distance relation is determined by the
  formula $\ell_{\rm luminocity}(z) = (1+z)^{2}d(z)$.
 The  factor $(1+z)^{2}$ comes  from the evolution of the
 angular size of the light cone of absorbed photons~\cite{039}.
 Since measurable distances in the conformal cosmology
 are the coordinate ones,
 we lose the factor $(1+z)^{-1}$
 that was in  the standard cosmology due to the expansion of the universe.
 Finally we obtain
 the redshift-luminosity distance relation
 \be\label{cldrr}
 \ell_{\rm luminocity}(z) = (1+z)^{2}d(z)=\frac{c}{H_0}
 \left[z+\frac{z^2}{2}\right]
 \ee
 as the consequence of  the  "inertial motion"
 of the universe along the geodesic line of the field {\it space}
 (i.e., the stiff state of dark energy with the most singular behaviour).

  It has been shown in paper~\cite{039} that this relation does not contradict
 the latest Supernova data~\cite{snov,riess,sn1997ff}.
 Among the CC models the pure stiff state of
dark energy gives the best
description  and it is equivalent to the
SC fit up to the distance of SN1997ff.


\subsection{Primordial element abundance}

 In the considered model of the
conformal cosmology the temperature is a constant. In the conformal
cosmology we have the mass history~\cite{039}
\be \label{mz} m_{\rm
era}{(z_{\rm era})}=\frac{m_{\rm era}(0)}{(1+z_{\rm era})}
\ee
with the constant temperature $T= 2.73 ~ {\rm K} = 2.35
 \times 10^{-13}$ GeV
where $m_{\rm era}(0) $ is characteristic energy (mass) of the
era of the universe evolution, which begins at the redshift $z_{\rm era}$.

 Eq. (\ref{mz}) has an important consequence that all
 physical processes, which concern the chemical composition of the
 universe and depend basically on the Boltzmann factors with the
 argument $(m/T)$, cannot distinguish between the conformal
 cosmology $(\bar m/\bar T)$ in the stiff state with
 the square root dynamics~(\ref{evol}) (in the "relative" standard)
 \be
 \label{ccd}
 a (\eta)= \sqrt{1+2H_0(\eta-\eta_0)}
 \ee
 and the  FRW cosmology in the radiation state with
 the same square root dynamics (in the "absolute" standard)
  due to the relations
 \be\nonumber
 \frac{\bar m(z)}{\bar T(0)}=\frac{m(z=0)}{(1+z)T(z=0)}=\frac{m(0)}{T(z)}~.
 \ee
 From this formula it is clear that the
 $z$-history of masses with invariant temperatures in the stiff
 state of conformal cosmology is equivalent to the $z$-history of
 temperatures with invariant masses in the radiation stage of the
 standard cosmology. We expect, therefore, that the conformal
 cosmology allows us to keep the scenarios developed in the
 standard cosmology in the radiation stage for, e.g. the
 neutron-proton ratio, primordial element abundance, and
 the appearance of CMB radiation with the temperature $2.7 K$.

Instead of the z-dependence of the temperature in
an {\it expanding universe} with  constant masses in the standard
cosmology, in conformal cosmology, we have the z-history of masses
in a {\it static universe} with an almost constant temperature of
the photon background (with the same argument of the Boltzmann
factors).

 Thus, the "relative" cosmology~\cite{039} leads to the simplest
 Cold Universe Scenario
 with the fundamental parameter of the CMB temperature $2.7 K$. In this
 Cold Scenario the single   stiff state  (treated
 as a primordial inertial motion along a geodesic line of the field
 space) describes
 two last eras: 2) the chemical evolution, and  3) the present-day
 stage. One can suppose that the same inertial
 motion~(\ref{gsp}),~(\ref{evol}), and ~(\ref{pr})
 describes also the primordial era of creation of the universe and creation
 of matter at the beginning of the universe. The description of the cosmological
 creation of matter
  requires a complete and consistent solution of
 problems of cosmic singularity $a=0$, cosmic initial data, and a positive
 arrow of the geometric time at the level of the quantum cosmic
 mechanics. We consider the solution of all these problems using
 the simplest case of the inertial motion.

\section{Hamiltonian Description of Quantum Relativistic Universe}

 \subsection{Hamiltonian formalism of cosmic inertial motion}

 The action of a relativistic universe~(\ref{1unco2})
 in terms of the canonical momentum~(\ref{canun})
 of the field evolution parameter $X_0$
  takes the
 form~\cite{M,MR}
 \be\label{ryan}
  S_{universe}
 =\int\limits_{x^{0}_{1}}^{x^{0}_{2}} dx^{0}\left[-P_0\partial_0X_{0}+
 \gamma e_0\left( \frac{P_0^2}{4\gamma^2}-H_0^2\right)
 \right]~.
 \ee
   The  Hamiltonian form~(\ref{ryan}) of the action~(\ref{1unco2})
 is well-known as the Dirac generalized mechanics~\cite{d2}
 of relativistic systems, in particular, a relativistic particle~(\ref{SR}).
 In this action~(\ref{ryan}) one of components of the metric in the
 Einstein theory
  plays the role of the time-like "coordinate of the Minkowskian space-time".
 Therefore, we call the variable $X_{0}$ the {\it world field time}.
 Whereas, the Misner {\it geometric} interval~(\ref{uncol0})
 \be \label{gi}
 d\Omega=e_0dx^0~~~~\Rightarrow~~~~
 \Omega=\int\limits_{x^0_1 }^{x^0_2 } dx^0 e_0(x^0)
 \ee
  is connected by eq.~(\ref{gsX})
  with the conformal time~(\ref{ct}) measured by the
  watch of an observer. This {\it geometric} interval
 is invariant with respect to reparametrizations of the coordinate
 evolution parameter
 \be \label{repa}
 x^0~\rightarrow~\tilde{x}^0= \tilde{x}^0(x^0)~,
 ~~~e_0~\rightarrow~ \tilde{e}_0~.
 \ee
 Classical equations of the generalized Hamiltonian system~(\ref{ryan})
 split into the equation of motion
 \be \label{em}
  \frac{d X_{0}}{e_0d x^0}=\frac{d X_{0}}{d \Omega}=\frac{P_{0}}{2\gamma},
  \ee
  $$~~~~~~~
  \frac{d P_{0}}{e_0d x^0 }=\frac{d P_{0}}{d \Omega}=0
 $$
 and the energy constraint
 \be \label{ec}
 \frac{P_{0}^2}{4\gamma^{2}}-H_0^2=0~.
 \ee
  One can see that solutions of these equations are expressed
  in terms of the invariant geometric interval~(\ref{gi})
  \be \label{RT}
 X_{0}(\Omega)=X_{0}(0)+\frac{P_{0}}{2\gamma}\Omega~,
 \ee
 \be \label{ecs}
 P_{0}=\pm 2\gamma H_0~.
 \ee
 We have seen above that these solutions can describe the classical
 evolution of the universe. However,
 these solutions do not allow us to determine the dependence
 of the lapse-function $e_0(x^0)$ on  the coordinate
 parameter $x^0$. There is  ambiguity in the lapse function $e_0(x^0)$,
 as it can be an arbitrary function.
 On the other hand, we need the lapse function as the variation
 of the action with respect to it
 leads to the energy constraint~(\ref{ecs}).
 The problem arises whether we should fix $e_0(x^0)$ or not.
 Here, we face  different mathematical and physical
 statements of the problem of the description of a constrained
 system.

 \subsection{Creation of quantum universe in {\it world field space}}

 The {\bf mathematical statement of the problem is
  to give a logically consistent description}
  of the considered relativistic system in its classical
  and quantum versions. This description includes
   solutions to all functions.
 If there is ambiguity, we have to fix it, for example,
 \be \label{necs}
 e_0(x^0)=1
 \ee
 In this case, $x^0$ is identified with the measurable time; and
 its Hamiltonian (that coincides with the constraint~(\ref{ec})),
  with a physical Hamiltonian.
 It seems that the problem of the classical description is
 completely solved.
 The Hamiltonian of evolution with respect to
 the "time" $x^0$ is identified with the constraint~(\ref{ec}).
 In the corresponding quantum theory (where
 $\hat P_{0}=id/dX_{0}$ is the operator),
 the wave function of the universe  satisfies
 the quantum version of the constraint~(\ref{ec})
 \be\label{wdw}
 \left(\frac{\hat P_{0}^2}{4\gamma^{2}}-H_0^2\right)\Psi_{wdw}(X_0)=0~,
 \ee
 This quantization
 for a universe is well known as the Wheeler-DeWitt (WDW) one.
  The WDW wave function as the amplitude of transition from  an initial
 scale factor $a_I=e^{X_{0I}}$ to a running scale factor $a_0=e^{X_{0}}$
 is decomposed over eigenvalues $P_0=\pm 2\gamma H_0$
  \bea\label{srw1}
 \Psi_{\rm wdw}(X_{0}|X_{0I})= A^+\Psi_{+}(X_{0}|X_{0I})
 \theta(X_{0}-X_{0I})
\nonumber\\
+A^-\Psi_{-}(X_{0}|X_{0I})\theta(X_{0I}-X_{0}),
 \eea
 where the wave functions $\Psi_{\pm}$  satisfy the equations
  \be\label{wdw1}
 \pm \frac{d}{idX_{0}}\Psi_{\pm}(X_{0}|X_{0I})=H_0\Psi_{\pm}(X_{0}|X_{0I})~.
 \ee
 Solutions  of these equations
  \be \label{srnw}
 \Psi_{\pm}(X_{0}|X_{0I}) ={\theta(\pm P_0)}
 \exp\{-iP_{0}(X_{0}-X_{0I})\}~,
  \ee
  $$
 P_0=\pm 2\gamma H_0
 $$
depend on the cosmic initial data $a_I=e^{X_{0I}}$
 (\ref{ecs})
 considered as an input parameter of the theory.
 The coefficient $A^{+}$ in the second
 quantization
 $$
 \left[A^{-},A^{+}\right]=1
 $$
 is treated as an operator of the creation of a universe with positive energy;
 and the coefficient $A^{-}$, as an operator of annihilation  of
 an anti-universe also with positive energy.
 The physical states are formed by the action of these operators on vacuum
 $<0|,|0>$ in the form of out-state (~$|1>=A^+|0>$~) with
 positive frequencies
 and in-state (~$<1|=<0|A^-$~) with
 negative frequencies.
 This treatment means that positive frequencies propagate forward
(${X_0}>{X_{0I}}$);
 and negative frequencies, backward (${X_0}>{X_{0I}}$) so that
 the negative values of energy are excluded from the spectrum
 to provide the stability of a quantum system~\cite{bww}.

  The causal Green function is defined
  as the probability to find a universe  at the moment $X_0$:

 \bea \label{caus}
 G^c(X_0)=&& <0|T(\Psi(X_0)\Psi(0))|0> \nonumber\\
&&
 \equiv
 G_+(X_0)\theta(X_0)+G_-(X_0)\theta(-X_0)\nonumber\\
&&=
  \frac{i}{2\pi} \frac{\exp(-i P_0 X_0)}{P_0^2-4\gamma^2 H_0^2-i\epsilon},
\eea
 where $G_+(X_0)=G_-(-X_0)$ is the "commutative" Green function~\cite{bww}
 \be \label{FIp}
 G_{+}(X_0)=
 \exp(-i P_0 X_0)\delta(P^2-4\gamma^2 H_0^2)\theta(P_0)~.
 \ee
 We see that the field variable $X_0$ plays the role of the evolution
 parameter that is the analogue of the time in the rest frame in Special
 Relativity.
 In both the cases Cosmic Relativity and Special Relativity
  the invariance of  actions with respect to
 reparametrizations of the coordinate time
 means that one of dynamic variables becomes a parameter of
 evolution. Recall that this identification was the main feature of the
 approach of Poincare~\cite{poi} and Einstein~\cite{ein} to the solution
 of the problem of energy of a relativistic particle in Special Relativity.
  Their result  is well-known in the form of $E=mc^2$ as
  the basis of nuclear power engineering.

  The wave function~(\ref{wdw1}) in the gauge
 $e_0=1$ does not contain the most interesting physical information
 about the evolution of the universe discussed  in Section 2.9.
 This means that the quantum description of the universe in gauge
 $e_0=1$ is not complete. How can  the evolution of
 the quantum universe with respect to the time measured by an observer
 be described?

 \subsection{Incorporation of time into quantum universe}

  To give the total description
  of the quantum universe, we
  create the time by the Levi-Civita
   canonical transformation~\cite{lc,sh,gkp,pp} of
   the {\it world field space}
 into {\it world geometric space}
\be\label{lc}
 (P_{0}, X_{0}) \Rightarrow\, (\Pi_{0}, Q_{0})
 \ee
 to new variables ($\Pi_{0},Q_{0}$) for which one of equations identifies
 new scale factor $Q_0$  with the geometric interval $\Omega$.

 This transformation~\cite{lc} is chosen so that to
 convert the constraint into the new momentum
 \be \label{levi}
 \Pi_0= \frac{P_{0}^2}{4\gamma}~ ,~~~~~~~~~~~~~~~~
 Q_0=X_0\frac{2\gamma}{P_0}~.
 \ee
 After transformation~(\ref{levi}), the action~(\ref{ryan}) takes the form
 \be \label{SRlc}
 S=\int\limits_{x^0_1}^{x^0_2} dx^0
 \left[
  -\Pi_{0}\frac{dQ_{0}}{dx^0}-
 e_0(-\Pi_0+ \gamma {H^{2}_{0}} )-\frac{d}{dx^0}s^{lc}
 \right]~,
 \ee
  where $s^{lc}=(Q_0 \Pi_0)$ is the generating functional of
  the canonical transformation.

 We can check that the equation of motion for the momentum $\Pi_0$
 \be\label{gecc}
 \frac{\delta S}{\delta \Pi_0}=0~\Rightarrow~\,dQ_0=e_0dx^0\equiv d\Omega.
 \ee
 identifies  the dynamic evolution parameter $Q_0$ with the
 geometric interval $\Omega$.

 The solution of  the equation of $e_0$ (i.e., the constraint)
 \be \label{gec}
 -\Pi_0+\gamma {H^{2}_{0}}=0~\Rightarrow~ \Pi_0=\gamma {H^{2}_{0}}
 \ee
 determines a new Hamiltonian
 of evolution with respect to the new dynamic evolution parameter $\Omega$.
 This Hamiltonian is not equal to zero in  contrast to
 the gauge-fixing way of the description of a universe.

 The substitution of all geometric solutions~(\ref{gec}),~(\ref{gecc})
\be
Q_0=\Omega,~~\Pi_0=\gamma {H^{2}_{0}}
\ee
into the inverted Levi-Civita transformation~(\ref{ivel})
\be \label{ivel}
 P_0=\pm 2\sqrt{\Pi_{0}\gamma}~,~~~~~~
 X_0=\pm Q_0{\sqrt{\frac{\Pi_{0}}{\gamma}}}
 \ee
 leads to the
conventional  relativistic solution for the field system~(\ref{RT})
\be \label{line0}
P_0=\pm 2\gamma{H_0}~,
\ee
$$
X_0(\Omega)= X_0(0)+\Omega\frac{P_0}{2\gamma}~.
$$
The quantization $-i[\Pi,Q_0]=1$ means that instead of constraint~(\ref{gec})
we have the Schr\"odinger equation for the wave function
\be \label{geom0}
\frac{d}{id\Omega}\Psi_{lc}(\Omega)=\gamma {H^2_{0}}\Psi_{lc}(\Omega),~
\ee
$$
\Psi_{lc}(\Omega)=  \exp(i\Omega{\gamma H^2_{0}})\exp(is^{lc})=
\exp(+i2\Omega{\gamma H^2_{0}})
$$
that contains only one eigenvalue $\gamma {H^2_{0}}$.
We see that there are differences between the {\it field} (Poincare-Einstein)
and  {\it geometric} (Levi-Civita) descriptions.
The field evolution parameter is given in the whole region
$-\infty < X_0 < +\infty$,
whereas the geometric one is only positive $0 <\Omega< +\infty$, as it
follows from the properties of the causal Green function~(\ref{caus})
after the Levi-Civita transformation~(\ref{levi})
$$
G^c(Q_{0})=\frac{i}{2\pi}\int\limits_{-\infty}^{+\infty}d\Pi_0
\frac{\exp(iQ_{0}\Pi_{0})}{(\Pi_0-{\gamma H^2_{0}}-i\epsilon)}=
$$
$$
=\theta(\Omega)\exp(iQ_{0}\gamma H^2_{0}),~~~~~~Q_0=\Omega~.
$$
Two solutions of the constraint (a universe and an antiuniverse)
 in the {\it field space}
 correspond to a single solution in the {\it geometric space}.

 For the causal convention~(\ref{srw1}), the {\it geometric time}
  $\Omega$ in classical solutions~(\ref{RT})
 \be \label{at}
 \Omega({X_0},X_{0I})=
 \pm \frac{1}{H_0}({X_0}-X_{0I}) \geq 0
 \ee
  is always positive as a consequence of
 the stability of the corresponding quantum system.
 For an Einstein observer,  the negative time
 does not exist.

 Thus, the reparametrization-invariant content of the equations
 of motion of a
 relativistic universe in terms of the geometric interval is covered
 by two  systems: the field and geometric ones.
 The field  system  describes the secondary
 quantization and the derivation of the causal Green function that
 determines the arrow of the geometric interval ($\Omega$).
 At the same time,  the geometric set of variables  includes
 the geometric interval $(Q_0=\Omega)$ into the number of measurable
 quantities, and it gives us a second wave function of
 a relativistic universe observed in the {\it world geometric space}.

 The relations $X_{0}(Q_{0})$ between  these two wave functions
 in the form of the Levi-Civita canonical transformation are
 treated as  pure relativistic effects.
  These relativistic
 effects could not be described by a single Newton-like system.
 Any gauge of the type  $e_0=1$ is an attempt to reduce the
 relativistic system to a single Newton-like system. This gauge
 violates the reparametrization invariance and loses part of physical
 information, in particular,  the positive arrow
 of the geometric interval, the initial geometric data, and
 the fact of the origin of the time measured by the watch of
 a observer.
 They are just the cardinal problems of the modern
 cosmology: the positive arrow of the time, its origin, and the
 cosmic initial data.

 \subsection{Solution of the problem of cosmic singularity
 in quantum universe}

 To discuss the problem of cosmic singularity,
 we reconsider the problem of  evolution  of the inertial universe
  in the relative  space-time
 \bea\label{mi}
 d\bar s^2=(\bar N_0(x^0) dx^0)^2-\sum\limits_{i=1 }^{3 }(dx^i)^2~,\nonumber\\
\bar N_0(x^0) dx^0\equiv d\eta
 \eea
 in terms of the running Planck mass
 \be\label{run}
 \vh(x^0)=\vh_0a(x^0)=\vh_0e^{X_0(x^0)}~.
 \ee
  In this case, the action~(\ref{ryan}) takes the form
 \be \label{F}
 S^{F}=\int\limits_{x^0_1 }^{x^0_2 }d x^0
 \left[ -P_{\vh}\dot \vh + \bar N_0\left(
 \frac{P_{\vh}^2}{4 V_0}-\rho_I(\vh)V_0 \right)\right]~,
 \ee
 where $\rho_I(\vh)$ is given by  the "inertial"
  density that plays the role of
  the dark energy and almost coincides with the critical density
 at the present-day time
 \bea \label{Qi}
 \rho_I(\vh)|_{\vh=\vh_0}&=&\frac{H^2_{0}\vh_0^4}{\vh^2(\eta)}|_{\eta=\eta_0}
 =  H_0^2\vh_0^2\nonumber\\
&&
 \equiv H_0^2 M^2_{\rm Planck}\frac{3}{8\pi}\equiv \rho_{cr}~.
 \eea
 The equation of motion in the stiff state
 $$
 \frac{d\vh}{N_0dx^0}\equiv \vh'=\frac{P_{\vh}}{2V_0}=
 \pm \frac{H_0\vh_0^2}{\vh}~\Rightarrow~(\vh^2)''=0
 $$
 reproduces the square root dynamics~(\ref{evol})
 \be
 \label{ccd1}
 \vh (\eta)= \vh_I\sqrt{1+2H_I\eta}
 \ee
 that describes SN data, as we have seen before in Section 2.9.
 The universe at the beginning of the evolution $\eta=0$
 is characterized by two vacuum data: a primordial  value of
 the dynamic Planck mass
 and the primordial Hubble parameter
 \be\label{zero1}
 \vh(\eta=0)=\vh_I,~~~~~~~~~~~~~~~H(\eta=0)=H_{I}~.
 \ee
 These primordial data are
 connected with the present-day values of the dynamic Planck mass
  $\vh_0$  and the Hubble parameter $H_0$ by
 the integral of motion
 \be \label{iom}
 \vh^2({\eta}) H(\eta)=\vh_I^2 H_I=\vh_0^2 H_0={\rm constant}~.
 \ee

 The quantization of  the dynamic Planck mass $i[\vh,P_{\vh}]=\hbar$ leads
 to a quantum version of the energy constraint well-known
 as the Wheeler-DeWitt equation
 $$
 \left[
 -\frac{P_{\vh}^2}{4 V_0}+\rho_I(\vh)V_0 \right]\Psi_{wdw}=0~.
 $$
 Two possible values of the
 momentum $P_{\vh}=\pm 2V_0\sqrt{\rho_I(\vh)}=\pm 2\gamma H_0/\vh$
 (where $\gamma=V_0\vh^2_0$)
 correspond to two Wheeler-DeWitt wave functions
 \bea\label{WDW}
 \Psi_{ wdw}(\vh_I,\vh_0)&=&
 A^+ e^{iS_{+}(\vh_I,\vh_0)}\theta (\vh_0-\vh_I)~~~~~~~~~~~~\nonumber\\
&&
 +A^-e^{iS_{-}(\vh_I,\vh_0)}\theta (\vh_I-\vh_0)~,
 \eea
 where $S_{\pm}(\vh_I,\vh_0)$ coincides with the
 constrained action~(\ref{F})
 \be\label{Fq}
 S^{(F)}(\rm constraint)=S_{\pm}(\vh_I,\vh_0)=
 \mp 2\gamma H_0\log\frac{\vh_0}{\vh_I}
 \ee
 that keeps only the field variable.
 By  analogy with a relativistic particle we can treat
 the coefficients $A^+,A^-$ of the decomposition as the operator
 of creation of the universe and the operator of annihilation.
 The singularity $\vh_0 \to 0$ is contained in the second solution.

 Thus, the problem of singularity is solved in quantum theory
 by the nonzero initial data~(\ref{zero1})
 for the solution that corresponds to creation of the universe.

 The WDW wave function~(\ref{WDW}) is not complete as it
 loses  the dependence of masses on the time~(\ref{ccd1}).
  The complete description of the quantum relativistic universe
 is possible by two wave functions
 in two sets of variables: the {\it field system} and
 the {\it geometric} one.

 \subsection{Description of SN data in the Quantum Universe}

  To find the second geometric system, we use
   the Levi-Civita canonical transformation~\cite{pp,bpp}
 \be\label{LC}
 \left[P_{\vh},~~\vh\right]
 ~~\Rightarrow~~\left[\Pi_0,~~Q_0\right]
 \ee
 \be \label{LC1}
 \vh=\vh_I\exp\left\{\pm{\frac{ Q_0}{\vh_I}\sqrt{\frac{ \Pi_0}{V_0}}}\right\},
 \ee
 $$
 P_{\vh}= \pm \sqrt{\Pi_0V_0}
 \exp\left\{\mp{\frac{ Q_0}{\vh_I}\sqrt{\frac{ \Pi_0}{V_0}}}\right\},
 $$

 $$
 \bar N_0=e_0\exp\left\{\pm {\frac{2Q_0}{\vh_I}
 \sqrt{\frac{ \Pi_0}{V_0}}}\right\}
 $$
 that incorporates  geometric interval into
 the set of the geometric variables. We called
 this set the  geometric system. In terms of the new geometric variables
 the action of the relativistic inertial universe takes the form~(\ref{SRlc}).

 Finally, we get the set of the geometric variables
 for which the energy constraint coincides with
 the momentum $\Pi_0=\gamma H^2_{0}$,
 and the equation for this momentum $\Pi_0$
  points out that the new variable
 coincides with the definition of the geometric interval
\be\label{pi}
 dQ_0=e_0dx^0=\frac{d\eta }{1+2H_{I}\eta}
\ee
 leading to
$$
 Q_0(\eta)=Q_I+\frac{1}{2H_I}\log(1+2H_I\eta)~.
$$
 The corresponding wave function satisfies equation
 $\Pi_0\Psi_G(Q_0)=\gamma H^2_{0}\Psi_G(Q_0)$. The solution
 of this equation takes the form
 \be\label{LCw}
 \Psi_G[Q_0(\eta)]=\exp\left\{ -i2Q_0(\eta)\gamma H^2_{0}\right\}~.
 \ee
 The evolution of the universe~(\ref{ccd1}) is defined as
 a relation between the spectral parameter $\vh$ of the WDW
 wave function~(\ref{WDW})
 of the universe in the {\it world field space} and
 the  spectral parameter $Q_0(\eta)$ of the wave function~(\ref{LCw}) of
 the same universe in the {\it world geometric space}.
 This relation (described by the Levi-Civita transformations~(\ref{LC1}))
 gives us the dynamic status of the Hubble
 law  in the quantum universe as  a pure relativistic effect.

 In the considered case of the inertial motion in the coset A(4)/L,
  this Hubble law  in the quantum universe~(\ref{ccd1}) is
 compatible with the SN data.

 Due to the Levi-Civita  relation the causal quantization
 of the field spectral parameter $\vh$ gives  the region of
 definition of the geometric spectral parameter $\eta$ as
 the positive arrow of the time
 \be\label{+}
 \eta_0-\eta_I\geq 0 ~~~~(\vh_0 \geq \vh_I)~~~ P_{\vh}\geq 0~,
 \ee
 $$
~\eta_0-\eta_I\geq 0 ~~~~(\vh_I\geq \vh_0)~~~P_{\vh}\leq 0~.
 $$
 Thus, the positive arrow of the time and its beginning can be considered
 as the evidences for the quantum nature of the universe.

 The relativistic gauge-invariant description
 also solves  the problem of positive energy in quantum theory
 of gravitation as the negative sign in~(\ref{Fq}) corresponds to
 annihilation of the universe. If we  lived in the created universe,
 we should choose the positive sign.

 We can see that the problems of cosmic initial data and positive
 arrow of time cannot be solved by the standard description
 of quantum universe by the single WDW wave function.

 \section{Creation of matter from vacuum}

\subsection{Statement of the problem}

Recall that in the inflational models~\cite{linde} it is proposed that
from the very beginning the universe is a hot fireball of massless particles
that undergo a set of phase transitions. However, the origin of particles
is an open question as the isotropic evolution of the
universe cannot create massless particles. Nowadays, it is evident that
the problem of the cosmological creation of
matter from vacuum is beyond the scope of the inflational models.

Here we try to explain the cosmological creation of particles
 from vacuum in the regime of  inertial motion of the universe
 along geodesic in the coset
 A(4)/L in the framework of the conformal cosmology.
 There,  the cosmic evolution
in the Standard Model (SM) $S_{\rm SM}[F|M_{\rm Higgs}]$
is separated by the conformal
transformation $S_{\rm SM}[\bar F|a(\eta)M_{\rm Higgs}]$. In this
case, the spontaneous SU(2) symmetry breaking
and the vacuum expectation value of the
{\it relative} Higgs field $<\bar \Phi>=a(\eta)< \Phi>$ is determined by the
cosmic dynamics of the scale factor with the nonzero initial data
 described before in the regime of
the inertial motion in the coset $A(4)/L$.

The SN data and the chemical evolution of matter show evidence of
this regime in the origin of the universe.
In this case, using the SM perturbation theory we should  explain
not only the cosmological creation of particles of observational
matter from vacuum, but a primordial origin of the
temperature of these particles.

There are arguments in the favour of that conformal cosmology can explain a
cosmological creation of matter from vacuum in the regime of isotropic
evolution of the universe as creation of vector bosons due to their mass
singularity~\cite{hpp,sf}. The first estimations of this effect were made
in~\cite{114,yaf} in the conformal invariant models. Here, we present the
theoretical foundation of the creation vector bosons in the Standard Model in
the regime of the inertial motion in the coset $A(4)/L$.

 \subsection{Standard Model in Riemannian space}

 The matter field is included into the total action~(\ref{tot})
\be\label{tot1}
 S_{\rm tot}[\{F\}|\vh_0,c_0]=S_{\rm GR}[g|\vh_0]+
 S_{\rm SM}[g,\{f\}|c_0]
 \ee
  in the form of the Standard model of electroweak and strong interactions
 with the set of fields $\{F\}=g,\{f\}$ and the Higgs parameter $c_0$.
  The corresponding action  takes the form
 \be \label{Cw}
S_{\rm tot}=\int d^4x \sqrt{-g}\left[
{\it L}_{\Phi,\vh_0}+
{\it L}_{g}+{\it L}_{l}+
{\it L}_{l\Phi}+...\right],
\ee
where
\bea \label{LP}
{\it L}_{\Phi,\vh_0}=&&\frac{|\Phi|^2-\vh_0^2}{6}R \nonumber\\
&&
+D^{-}_{\mu}\Phi(D^{\mu,-}\Phi)^*
-\lambda\left( |\Phi|^{2}- c_0^{2} \right)^{2}
\eea
is the Lagrangian of  Higgs fields with
$$
D_{\mu}^{-}\Phi=(\partial_{\mu} -
\imath g \frac{\tau_{a}}{2}A_{\mu}-
\frac{\imath}{2}g^{\prime}B_{\mu})\Phi
$$
and
$$
\Phi = \left(\begin{array}{rr} \Phi_{+} \\ \Phi_{0}\end{array}\right),
$$
\be\label{Ph}
 |\Phi|^{2} = \Phi_{+}\Phi_{-}
 + \Phi_{0}\bar{\Phi}_{0};
\ee
\begin{eqnarray*}
{\it L}_{g}&=&-\frac{1}{4}\left(\partial_{\mu}A^{a}_{\nu} -
\partial_{\nu}A^{a}_{\mu}+g\varepsilon_{abc}A^{b}_{\mu}A^{c}_{\nu}
\right)^{2}\nonumber\\
&&-
\frac{1}{4}\left(\partial_{\mu}B_{\nu}
- \partial_{\nu} B_{\mu} \right)^{2}
\end{eqnarray*}
is the Lagrangians of gauge fields,
\bea \label{LL}
{\it L}_{l}&=&\imath \bar{L} \gamma^{\mu}D^{+}_\mu L
+ \imath \bar{e}_{R} \gamma^{\mu}\left(D^F_{\mu} +
\imath g^{\prime}B_{\mu}\right)e_{R}\nonumber\\
&&
 +\bar{\nu}_{R}\imath \gamma^{\mu}\partial_{\mu}\nu_R
\eea
and that of the leptons
\be \label{L1}
L = \left(
 \begin{array}{rr}
 \nu_{e,L} \\
  e_L \end{array}\right);~~e_R,\nu_{e,R}~;
\ee
$D^F$ is the Fock derivative,
$$D^{+}_\mu L =(D^F_{\mu} -
\imath g \frac{\tau_{a}}{2}A_{\mu}
+ \frac{\imath}{2}g^{\prime}B_{\mu})L.
$$

The Lagrangian describing mass terms of leptons including that of the
neutrino (if any) is

\bea \label{LVh}
{\it L}_{l\vh}&=& - y_{e}\left(e_{R}\Phi^{+}L
+ \bar{L}\Phi e_{R}\right)\nonumber\\
&&
- y_{\nu}\left(\bar{\nu}_{R}\Phi^{+}_{C}L
+ \bar{L}\Phi_{C}\nu_{R} \right),~~
\eea
where
\be
\Phi_{C} = \imath \tau_{2}\Phi^{+}_{C} =
\left(\begin{array}{rr}
 \Phi^{*}_{0} ~\\
 - \Phi_{-}\end{array}\right)
 ~~~~~~~~~~~~~~~~~~~~~~~~~~~~~~~~~~~~~~~~~
 \nonumber
 \ee
 and $y_f$ are dimensionless parameters. For simplicity, we omitted
 the strong interaction sector.

\subsection{Higgs effect in the inertial universe}

 The evolution of the universe in terms of the mass-scale factor
 \be\label{massca}
 \vh(x^0)=\vh_0 a(x^0),~~~~~~~~~~~~~~\vh_0=M_{\rm Planck}\sqrt{\frac{3}{8\pi}}
 \ee
 is separated  by the Lichnerowicz transformations~(\ref{lic})
 \bea\label{slic1}
 S_{\rm tot}[\{F\}|\vh_0, c_0]&=&
 S_{\rm tot}[ \{\bar F\}|\vh, y_0\vh]\nonumber\\
&&+
S_{\rm universe}[\vh,\bar N_0]~,
 \eea
 where $\bar{F}$ are observable fields and
\be\label{hassca}
 y_0=\frac{c_0}{\vh_0}
 \ee
 is the Higgs parameter in the Planck mass - units,
 \be\label{uncol1}
 S_{\rm universe}[\vh,\bar N_0]=- V_0
 \int\limits_{x^{0}_{1}}^{x^{0}_{2}} dx^{0}
 \left[ \frac{(\partial_{0}\vh)^2}{\bar N_0}+\bar N_0\rho_I(\vh) \right]
 \ee
 is the action of a collective inertial motion of the universe
 along the geodesic line of the field space with the
 primordial density
  \be\label{qun}
 \rho_I(\vh) \sim
 \frac{\vh_0^2 \rho_{\rm cr.}}{\vh^2}~,
 \ee
 is the primordial value of the mass-scale $\vh_I$, and
  $\bar N_{0}$ is
 the global lapse function that determines the invariant conformal time
 \be\label{ct1}
 d\eta=\bar N_{0}d x^{0}~.
 \ee
 It is just the time measured by the watch of an observer far from heavy masses
 where the Einstein relative interval takes a flat form
 \be \label{cst1}
 d\bar s^2=d\eta^2-dx_i^2~.
 \ee
 In this case, the Higgs potential
 \bea \nonumber
{\it L}_{\rm Higgs}= -\lambda\left( |\bar \Phi|^{2}
- y_0^{2}\vh^{2} \right)^{2}
\eea
describes the  Higgs effect of
the spontaneous SU(2)  symmetry breaking
 \bea \label{dps}
\frac{\partial {\it L}_{\rm Higgs}}{\partial |\bar \Phi|}
=0~\Rightarrow |\bar \Phi|_1 = 0,
\eea
$$
|\bar \Phi|_{2,3} = \pm {y_0}\vh\sim 10^{-17}\vh~.
$$
Two last stable solutions~(\ref{dps})
 form dynamic masses of elementary particle
in the units of the dynamic Planck mass $\vh^2=\vh_I^2(1+2H_0\eta)$,
in the inertial universe~(\ref{uncol1}).

The masses of elementary particles in the Lagrangian of SM
\bea \nonumber
 {\it L}=  y_e|\bar \Phi|  \bar e e+...,
\eea
remind the evolution of the universe.

\subsection{Perturbation theory}

The first step of the perturbation theory is to consider the independent
"free" fields in the linear approximation of their equations of motion.

 The perturbation theory in the relative field space~\cite{ps1}  begins from
the metric
\be \label{crin1}
 ds^2 = \frac{\vh^2(x^0)}{\vh_0^2}d \bar s^2~,
\ee
where the relative interval reads
\be \label{crin0}
d\bar s^2_0=d\eta^2 - [\delta_{ij}+2h_{ij}]dx^idx^j~,
\ee
where
$$d\eta = \bar N_0(x^0)dx^0~.
$$
We keep only independent local field variables
$h_{ii}=0,\partial_i h_{ij}=0$ which are  determined by
independent initial values. All nonphysical variables (for which the
initial values depend on other data) are excluded by the local constraint.

The substitution of the ansatz~(\ref{crin1}),~(\ref{crin0}) into the
action~(\ref{Cw}) leads to the action of free fields in terms of
physical variables~\cite{ps1,hpp}
\be \label{gradl}
S_{\rm tot}=\int\limits_{x^0_1 }^{x^0_2 }dx^0{\bar N_0}
\left[ -{V_0} \left(\frac{d{\vh}}{{\bar N_0}dx^0}\right)^2+{V_0}\rho_I(\vh)
+L\right],
\ee
 where $V_0$ is a finite spatial volume, and $L$ is the sum
 $ L=L_0+L_{\rm int}$ of the Lagrangian of interaction $L_{\rm int}$
 and the free one
\be \label{H0l}
L_0=
\int\limits_{V_0 } d^3x
\left({\cal L}_{\rm \chi} + {\cal L}^{\bot}_{\rm vec}+ {\cal
L}^{||}_{\rm vec}+{\cal L}_{\rm rad}+{\cal L}_{\rm s}+
{\cal L}_{\rm h}\right)
\ee
is the total Lagrangian of free fields.
In particular,
\be \label{Hig}
{\cal L}_{\rm \chi}=\left\{{{\chi}'}^2+
\chi_i\left[ \vec{\partial}^2-4\lambda(y_h\vh)^2\right] \chi_i\right\}
\ee
is the Lagrangian of a deviation of the modulus of the Higgs field
$|\bar{\Phi}|=y_0\vh+\chi$,

and
\begin{eqnarray}
{\cal L}_{\rm vec}^{\bot} = {1\over
2}\left\{  {v'}_i^{\bot}{}^2+
v_i^{\bot}\left[ \vec{\partial}^2-(y_v\vh)^2\right] v_i^{\bot}\right\}~,
\nonumber\\
{\cal L}_{\rm vec}^{||} = -\frac{(y_v\vh)^2}{2}
\left[ {v'}_i^{||} {1\over
\left(\vec{\partial}^2-(y_v\vh)^2\right)}
{v'}_i^{||}+v_i^{||2}\right]
\end{eqnarray}
are Lagrangians of the transverse (${\cal L}_{\rm vec}^{\bot}$)
and longitudinal (${\cal L}_{\rm vec}^{||}$) components of the
W- and Z- bosons~\cite{hpp,sf}.
The Lagrangian of the fermionic spinor fields is given by
\be \label{hml}
{\cal L}_{s}= \bar
\psi\left\{-y_s\vh- i{\gamma_0}\partial_{\eta}
+i\gamma_j\partial_j\right\}\psi~,
\ee
where the role of the masses
is played by the dynamic Planck mass $\vh $ multiplied by
dimensionless constants $y_{v,s}$; ${\cal L}_{\rm rad}$ is the
Lagrangian of massless fields (photons $\gamma$, neutrinos $\nu$)
with $y_{\gamma}=y_{\nu} =0$, and
\be \label{hol}
{\cal L}_{\rm h}=\frac{\vh^2}{24}
\left\{\left({ h'}_{ij}\right)^2
-(\partial_kh_{ij})^2\right\}~
\ee
is the Lagrangian of gravitons as weak transverse excitations of spatial
metric  $h_{ii}=0$ with a unit determinant
of the three-dimensional metric~$\partial_jh_{ji}=0$ (everywhere
$f'=df/[\bar N_0dx^0]$).

To find the evolution of all fields with respect to the
proper time $\eta$, we use the Hamiltonian form of
the action~(\ref{Cw}) in the approximation~(\ref{crin0})
\bea \label{grad}
S_{\rm tot}=\int\limits_{x^0_1 }^{x^0_2 }dx^0
 \left\{\int\limits_{V_0 }
d^3x \sum\limits_{F}P_F\partial_0 f
-P_{\vh}\partial_0 \vh \right.\nonumber\\
\left.\phantom{\int\limits_{V_0 }}-{\bar N_{0}}{V_0}\left[-\frac{P_{\vh}^2}{4V^2_0}
+\rho_I(\vh)+\rho_m(\vh,F,P_F)\right]\right\}~,
\eea
where the Hamiltonian density
 \be\label{F2}
V_0\rho_m(\vh,F,P_F)=V_0\rho_{(2)}(\vh,F,P_F)+ H_{\rm int}
\ee
is a sum of the
densities $\rho_{(2)}(\vh,F,P_F)$ of free
fields $F$ and their interactions.
 In particular, the  massive scalar Higgs field
  is  described by
 the Hamiltonian~\cite{ps1}
 \be\label{qu}
 H_{(2)} =V_0 \rho_{(2)}=\frac{1}{2}\sum\limits_{k }^{ }
 \left[P_{\chi,k}^2+\omega_{\chi}^2(\vh,k) \chi_k^2\right]~,
  \ee
 where $\omega_{\chi}^2(\vh,k) =\sqrt{k^2+y_\chi^2\vh^2}$.

The variation of the action with respect to
the homogeneous lapse-function $\bar N_0$ yields the energy constraint
\be
\label{d56}
\frac{\delta W_{0}}{\delta N_0}=0~
\Rightarrow\,
\vh'^2=\rho_I(\vh)+\rho_m(\vh)~.
\ee
Recall that we supposed that  $\rho_I(\vh)\gg \rho_m(\vh)$.

\subsection{Hamiltonian of Quantum Field Theory in  {\it world field space}}

 The substitution of the solution of  the
 energy constraint
 \be\label{Fc}
 \frac{P_{\vh}^2}{4 V_0}=V_0\rho_I+ H_m~\Rightarrow~
 P_{\vh}=\pm 2V_0\sqrt{\rho_I+ \rho_m}
 \ee
 into the action~(\ref{grad})
 leads to the Poincare-Einstein type actions with the field
 evolution parameter $\vh$
 \be\label{Fr1}
 S^{(F)}_{\pm}=\int\limits_{\vh_I }^{\vh_0 }d\vh
 \left\{\sum\limits_{F }^{ }P_F
 \frac{d F}{d\vh} \mp2V_0\sqrt{\rho_I+ \rho_m}
 \right\}~
 \ee
 that describes the evolution of "free"  massive fields with respect
 to the dynamic Planck mass.

  If the inertial density is greater than matter,
  $\rho_I(\vh)\gg \rho_m(\vh)$,
  the actions~(\ref{Fr1}) can be decomposed in a nonrelativistic form
 \be\label{Ft}
 2V_0\sqrt{\rho_I+ H_m/V_0}=2V_0\sqrt{\rho_I}+ \frac{H_m}{\sqrt{\rho_I}}+...~.
 \ee
  The substitution of~(\ref{Ft}) into~(\ref{Fr1}) leads to the sum
  $S^{(F)}_{\pm}=S^{(I)}_{\pm}+S^{(m)}_{\pm}$ of
  the  action of the inertial motion~(\ref{Fq}) $S^{(I)}_{\pm}$
  and the action of matter
  \be\label{Fm}
 S^{(m)}_{\pm}=
 \int\limits_{\vh_I }^{\vh_0 }d\vh
 \left\{\sum\limits_{F }^{ }P_F
 \frac{d F}{d\vh} \mp \frac{H_m}{\sqrt{\rho_I}}
 \right\}~.
 \ee
 The neglect of the "back-reaction" allows us to use
 the solution of the constraint $P_\vh=2V_0\sqrt{\rho_I(\vh)}$
 $$
  d\vh(\eta) =  \sqrt{\rho_I} d\eta
 $$
 In this case the action~(\ref{Fm}) takes a form of the sum of the
 action of the inertial motion and the one of the matter field in the
 form of a standard action in quantum field theory
 \be\label{Fm1}
 S^{(m)}_{\pm}=
 \int\limits_{\eta_I }^{\eta_0 }d\eta
 \left\{\sum\limits_{F }^{ }P_F
 \frac{d F}{d\eta} \mp {H_m}
 \right\}~.
 \ee
 Thus,  we  derived the ordinary action of QFT with
 the measurable conformal time $\eta$ without
 the concept of nonlocal energy~\cite{fp2}.

\subsection{Holomorphic representation and number of particles }

 The relativistic system is given in the world field space
 $(\vh,F_k)$, where $F_k$ are oscillators treated as  "particles".
  We  define "particles"  as holomorphic field variables
\be\label{Fm10}
   F(t,\vec x)=
\ee
$$
\sum\limits_{k}^{ }
\frac{C_F(\vh)e^{ik_ix_i}}{\sqrt{2V_0\omega_F(\vh,k)}}
\left( a_{\sigma}^+(-k,t)\epsilon_{\sigma}(-k)+
a_{\sigma}(k,t)\epsilon_{\sigma}(k)\right)~,
$$
where
\bea\nonumber
C_h(\vh) = \frac{\sqrt{12}}{\vh},~~~
C^{||}_{v}(\vh)=\frac{\omega_{v}}{y_{v}\vh},
\eea
$$
C_{\gamma}(\vh)=C_{s}(\vh)=C^{\bot}_{v}(\vh)=C_{\chi}(\vh)=1~.
$$
These variables are distinguished by that they
  diagonalize the energy density of "free" fields~(\ref{qu})
 \be\label{Fm100}
 {\hat \rho}_{(2)}(\vh)=
 \sum_{\varsigma }^{ }\omega_{F}(\vh,k){\hat N}_{\varsigma}~,
 \ee
where $\omega_{F}(\vh,k)=(k^2+y_F^2\vh^2)^{1/2}$ is the
one-particle energy; ${\hat N}_{\varsigma} =
\frac{1}{2}(a_{\varsigma}^{+}a_{\varsigma} +
a_{\varsigma}a_{\varsigma}^{+})$ is the number of particles;
$\varsigma$ includes momenta $k_i$, species
$F=h,\gamma,v,s$, and spins $\sigma$.
As everyone can see the longidutial vector paricles
have the mass singularity $m_{v} = y_{v} \vh$~\cite{hpp,sf}.

Diagonalization of the density~(\ref{Fm100}) can be treated as  a rigourous
definition of a ``particle'' in quantum field theory that is consistent with
 observational cosmology. Really, in cosmology we
 consider the universe
as a collection of ``particles'' with definite energies.

Just for observable ``particles'' the equations of motion
are not diagonal. In particular, after transformation~(\ref{Fm10})
 the canonical differential form in the action
\bea \nonumber
\int\limits_{V_0 }d^3x \sum\limits_{F}P_F\partial_0 F
=\sum_{\stackrel{\scriptstyle\varsigma~ =}
{\scriptstyle (k, F,\sigma)}}
\frac{\imath}{2}( a^{+}_{\varsigma} {\partial_0 a}_{\varsigma}
-a_{\varsigma}{\partial_0 a}^{+}_{\varsigma} )
\eea
$$
- \sum_{\varsigma}
\frac{\imath}{2}(a^{+}_{\varsigma}a^{+}_{\varsigma} -
a_{\varsigma}a_{\varsigma}) \partial_0 \Delta_{\varsigma}(\vh)
$$
acquires nondiagonal terms as sources of cosmic creation of particles.

The set of nondiagonal terms in SM is
\bea \nonumber
\Delta_{h}(\vh) = \log(\vh /\vh_{I}),
~~~\Delta^{{\bot}}_{v}(\vh) =\Delta_{\chi}(\vh) =
\frac{1}{2}\log(\omega_{v} / \omega_{I}),
\eea
$$
\Delta^{||}_{v}(\vh) =\Delta_{h}(\vh) - \Delta^{\bot}_{v}(\vh),
$$
where $\vh_I$ and $\omega_I$ are initial values.

 \subsection{Cosmological creation of vector bosons}

The number of created particles is calculated by diagonalization
of the equations of motion by the Bogoliubov transformation
\be \label{bogtr}
b_{\varsigma}=\cosh(r_{\varsigma})
e^{ i \theta_{\varsigma}}a_{\varsigma}
 + \imath\sinh(r_{\varsigma})
 e^{-i \theta_{\varsigma}}a_{\varsigma}^{+}~.
\ee

 These transformations play the role of the Levi-Civita
canonical transformation to the action-angle variables that
 give integrals of motion for the matter fields
 \be \label{LC21}
 \frac{d}{d\eta}b^+b=0~.
 \ee
 These vacuum cosmic initial data correspond
 to the state {\it nothing} $b|0>=0$.
\begin{figure}[ht]
 \vspace{2mm}
 \begin{center}
 \includegraphics[width=0.23\textwidth,clip]{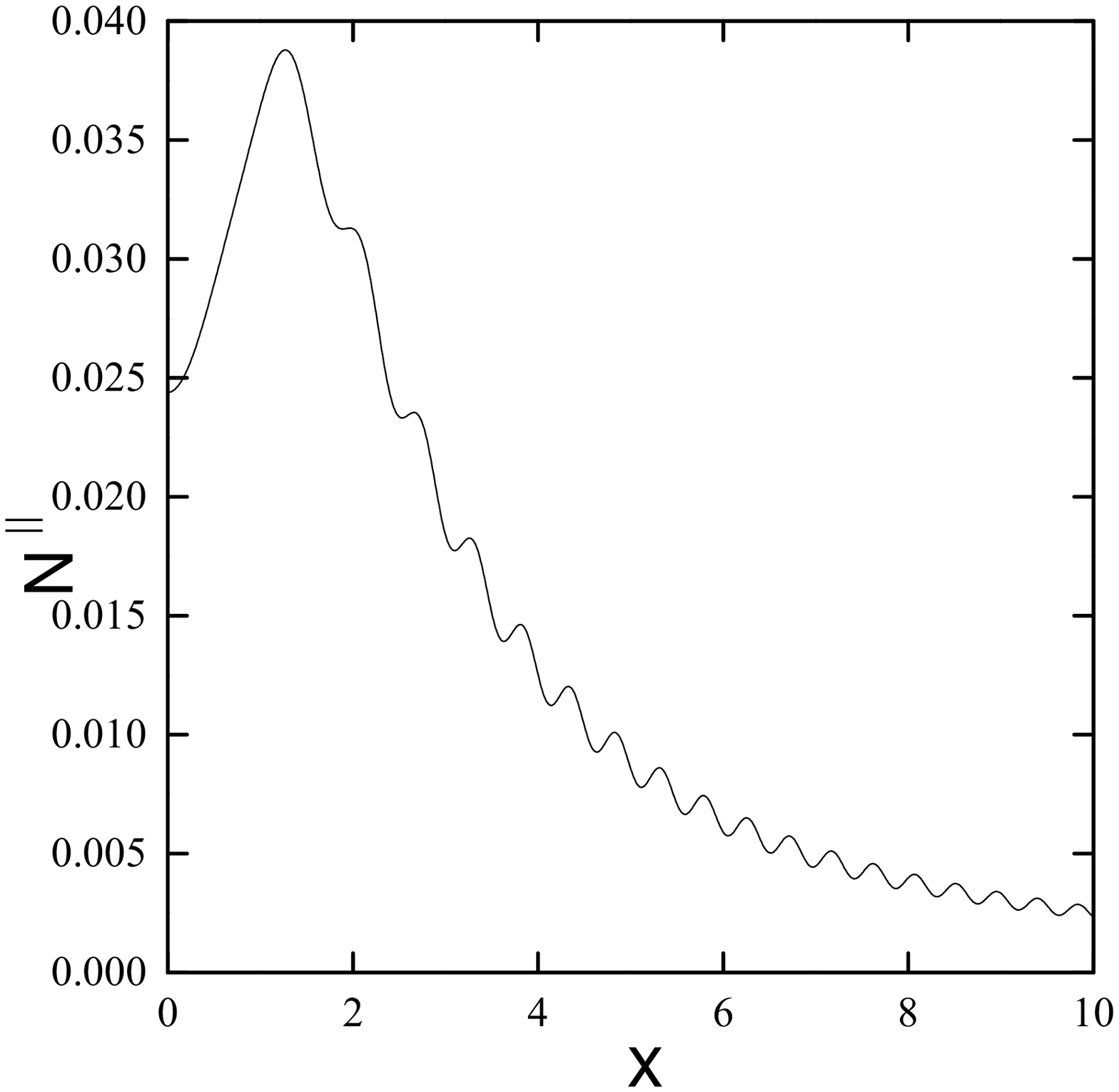}
 \includegraphics[width=0.23\textwidth,clip]{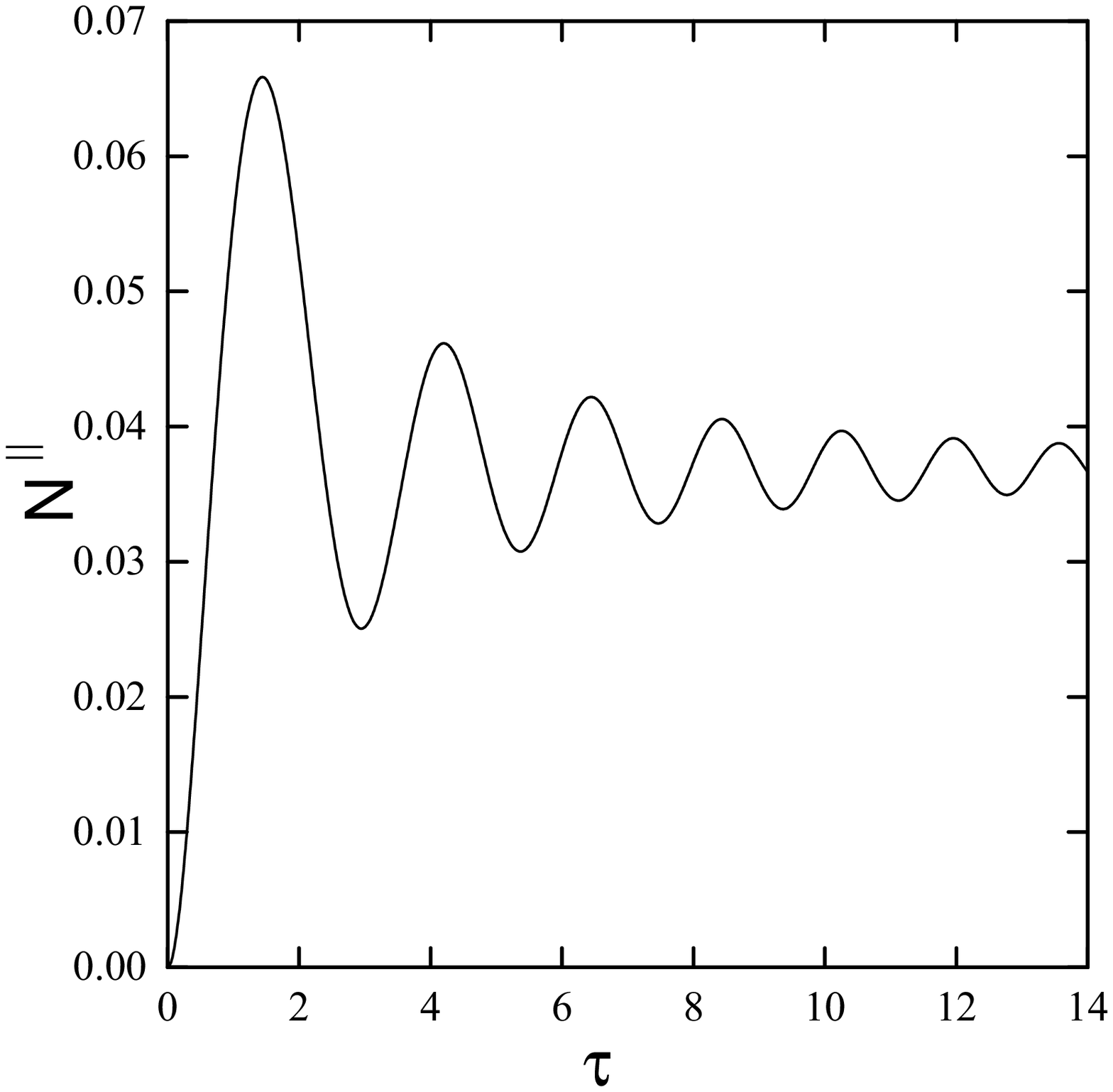}\\
 \includegraphics[width=0.23\textwidth,clip]{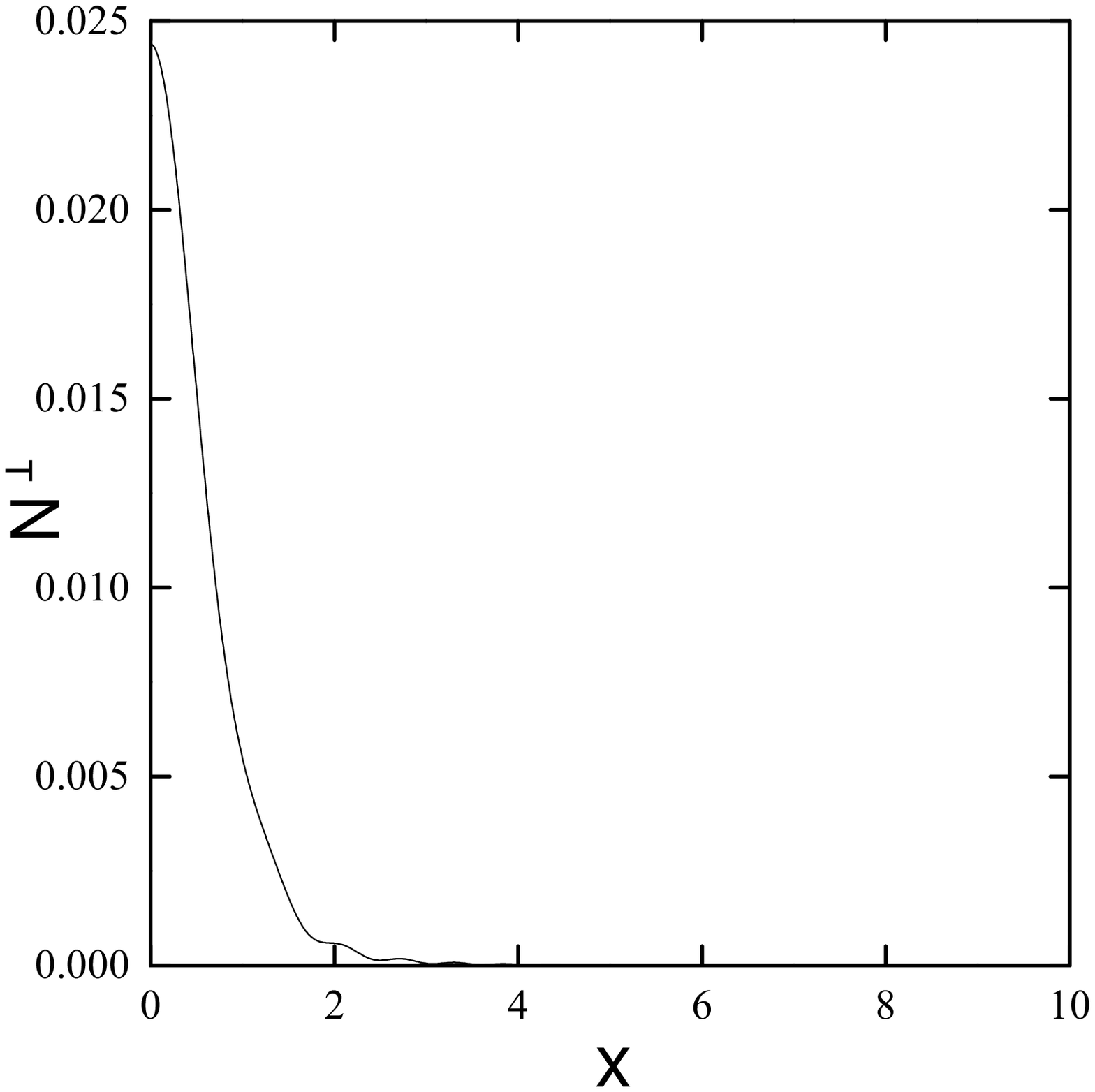}
 \includegraphics[width=0.23\textwidth,clip]{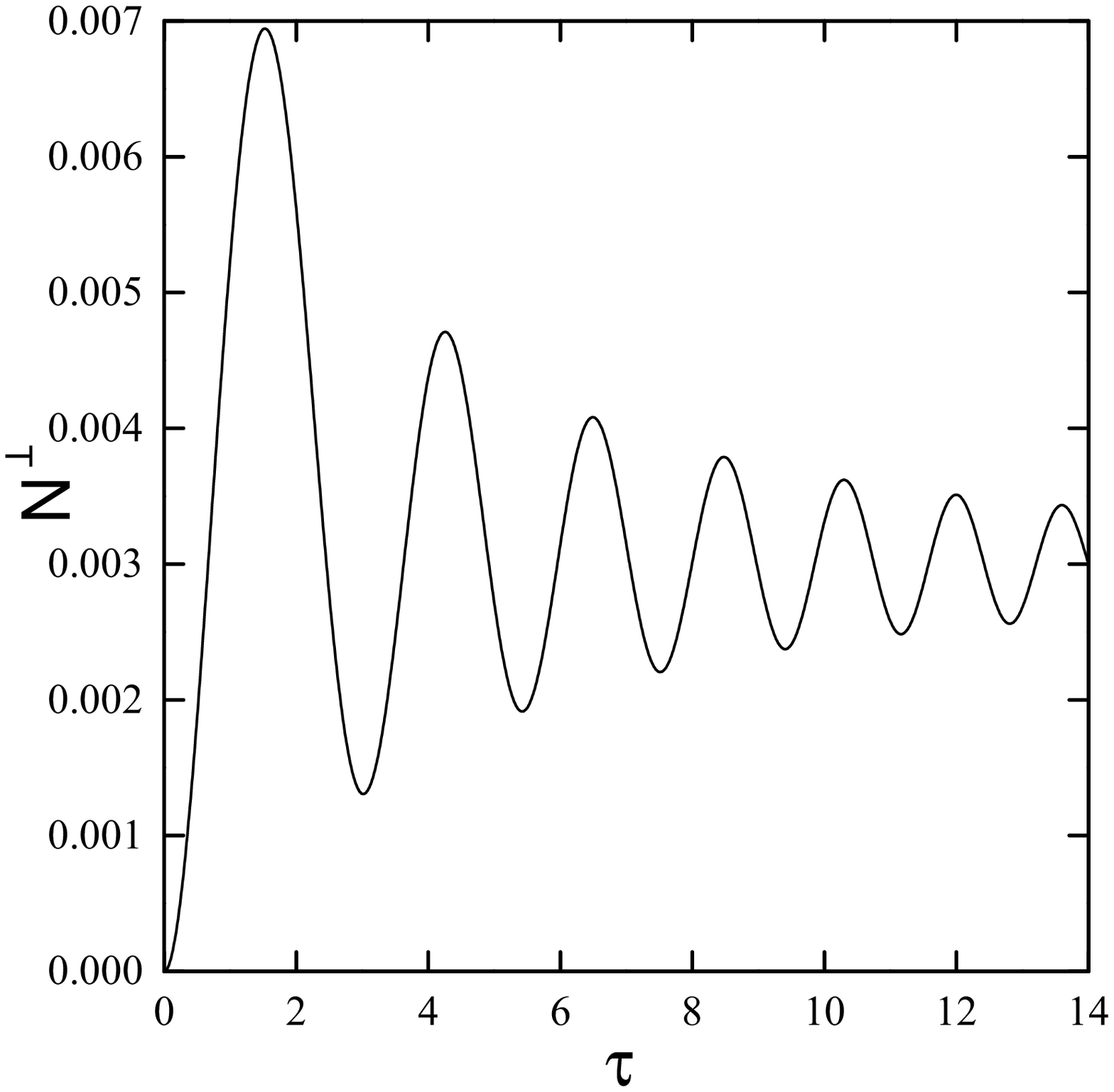}
\caption{
Time dependence  for the dimensionless momentum $x=k/H_I=1.25$ (left panels)
and momentum dependence  at the dimensionless lifetime $\tau=(\eta 2H_I)=14$
(right panels)
of the transverse (lower panels) and
longitudinal(upper panels)
components of the vector-boson distribution function.}
\label{fig2}
 \end{center}
\end{figure}

The equations for the  Bogoliubov coefficients
\bea
[\omega_{\varsigma} - \theta'_{\varsigma}] \sinh(2r_{\varsigma}) &=&
\Delta'_{\varsigma}\cos(2\theta_{\varsigma})\cosh(2r_{\varsigma}),\nonumber \\
r'_{\varsigma} &=& - \Delta'_{\varsigma}\sin(2\theta_{\varsigma})\nonumber
\eea
determine the number of particles
\be \label{Number}
{\it N}_{\varsigma}(\eta) =
{}_{sq} \langle 0|\hat N_{\varsigma}|0\rangle_{sq}-
{1}/{2}=\sinh^2 r_{\varsigma}(\eta)
\ee
created during the time $\eta$ from squeezed vacuum:
$b_{\varsigma}|0\rangle_{sq} = 0$
and the evolution of the density
$$\rho(\vh)=\vh'^{2} = \sum_{\varsigma} \omega_{\varsigma}(\vh)
{}_{sq}\langle 0|\hat N_{\varsigma}|0\rangle_{sq}~.
$$

The numerical solutions of the Bogoliubov equations
 for the time dependence of the vector boson distribution
functions ${\it N}_v^{||}(k,\eta)$ and ${\it N}_v^{\bot}(k,\eta)$ are given
in Fig.~\ref{fig2} (left panels) for the momentum $k=1.25 H_I$.
We  can see that the longitudinal function
is noticeably greater than the transverse one.
The momentum dependence of these functions at the beginning
of the universe is given on the right panels of Fig.~\ref{fig2}.
The upper panel shows us the intensive cosmological creation of
the longitudinal bosons in comparison with
the transverse ones.
This fact is in agreement with the mass singularity of
the longitudinal vector bosons discussed in~\cite{hpp,sf}.
One of the features of this intensive creation is a high momentum
tail of the momentum distribution of longitudinal bosons which leads
to a divergence of  the density of created particles defined
as~\cite{par}
\be\label{nb}
n_{v}(\eta)=\frac{1}{2\pi^2}
\int\limits_{0 }^{\infty } dk k^2
\left[ {\it N}_v^{||}(k,\eta) + 2{\it N}_v^{\bot}(k,\eta)\right]~.
\ee
The divergence is a defect of our approximation
where we neglected all interactions of vector bosons that form
the collision integral in the kinetic equation for the distribution
functions.

Our calculation of this density presented in Fig.~\ref{fig2}  signals that the
density~(\ref{nb}) is established very quickly in comparison with
the lifetime of bosons and in the equilibrium
there is a weak dependence of the density on the time (or z-factor).
This means that the initial Hubble parameter $H_I$  almost
coincides with the Hubble parameter at the point of saturation.

 \subsection{Towards the microscopic theory of CMB temperature}

 A new fact is the zero-mass singularity
of the longitudinal vector bosons~\cite{sf,hpp}
that leads to divergence of the number of created vector bosons
in the lowest order of  perturbation theory~\cite{114}
\be\label{Numberw}
n^{\|}=\frac{1}{2\pi^2}\int dk k^2 N^{\|}(k)\sim \infty~.
\ee
This divergence is the real origin of appearance
of temperature of the created matter in the steady universe
of the relative cosmology. This temperature belongs to
vector bosons.

 The concept of a {\it temperature} $T$
 for the matter fields in the relativistic region
 means that the distribution function
 $N = |\sinh^2 r|$ converts into the Boltzmann factor, and
 the particle density takes the form $n(T)\sim T^3$.
 This {\it temperature} is established
 due to the interaction cross-section
 $$
 \sigma_{\rm scat.} \sim \frac{1}{m^2_v(\eta_{\rm relax.})}
 $$
  during the  time of relaxation
 $$
 \eta_{\rm relax.}=\frac{1}{n(T)\sigma_{\rm scat.}}~.
 $$
 We can introduce the concept of {\it temperature} $T$
 if this  time of relaxation is less than the inverse
 Hubble parameter (i.e., the  universe is varying slowly than
 fields)
 $\eta_{\rm relax.}\leq 1/H(\eta_{\rm relax.})$.
 This means that the temperature T can be estimated by the integral of
 motion
$$
 n(T)\sigma_{\rm scat.} \geq H(\eta_{\rm relax.})~
 \Rightarrow
$$
 \be
 ~T_I\sim
 \left(m^2_v(\eta_{\rm relax.})H(\eta_{\rm relax.})\right)^{1/3}=
 \left(m^2_{v0}H_{0}\right)^{1/3}~,
 \ee
 where $m^2_{v0},H_0$ mean the present-day values.
 If we suppose that the CMB radiation is the product of the
 decay of the primordial vector bosons and its temperature $T_{CMB}=2.7 K$
 remembers the primordial temperature $T_I=T_{CMB}=2.7 K$,
 we can obtain the present-day mass of the primordial vector bosons
 \be
 m_{v0}=\left(\frac{T_{CMB}^3}{H_0}\right)^{1/2}\simeq 80 GeV \div 100 GeV~.
 \ee
 It is just the mass of $W,Z$ - vector bosons
 in the Standard Model of the electroweak interactions.
 Using this mass we can estimate also
  primordial values of the of Hubble parameter
 and Planck mass. They  are close to
 \be
 H_{I} = T_{CMB}= 2.7 K,~~
 \vh_I=\frac{\vh_0T_{CMB}^{1/2} }{H_0^{1/2}}=10 TeV~.
 \ee
 One can say that the CMB-temperature "remembers" the
 primordial mass of the inertial motion of the universe.

 These estimations are in agreement with the
 latest Supernova data~\cite{039,snov,riess,sn1997ff},
  the chemical evolution,
  the positive arrow of the
 geometric time, and the
 inertial density
 $$
 \rho_I(\vh)=\frac{\vh_I^4 T_{CMB}^2}{\vh^2} \equiv
 \frac{\vh_0^2 \rho_{\rm cr.}}{\vh^2}~.
 $$
 It is varying in the time $\eta$ from $10^{29}\rho_{\rm cr.}$
 to $\rho_{\rm cr.}$ during the time $\eta = 1/2H_0$.
This density is  greater than the one of created bosons
$$
\frac{\rho_I(\vh_I)}{T^4_{\rm CMB}}=\frac{\vh^2_I}{T^2_{\rm CMB}}=10^{-34}~.
$$
 These estimations are compatible
 with a direct calculation of the primordial creation of vector particles~\cite{114,yaf}.
 This creation leads to the baryon asymmetry of the universe
 as the left-current interaction of the primordial vector bosons
 in SM during their lifetime polarize the Dirac sea of fermions~\cite{039}.

\section{Discussion of the results}

 The universe was considered as one of ordinary physical
 objects described by differential equations of the General Relativity
 and Standard Model  given in the definite frame of reference
 (connected with the CMB radiation)
  with the zero initial data for all fields $<\bar F>_I=0$, excluding
  the scale factor
  $$a_I=0.3 \cdot 10^{-14}~,~~~~~~~(a'/a)_I=2.7 K$$
  and the relative Higgs field
  $$<\bar \Phi>_I=a_I< \Phi>_0~,$$
   where
  $< \Phi>_0$ is the present-day value.

 It is useful to remind also that
 modern quantum field theories cannot explain a measurement
 standard (identifying the theoretical quantities with the
 observational ones) and initial data including
  the background of infrared gravitation fields
 (like an undetected background of infrared photons which should be taken into
  account for the consistent theoretical description of
  the higher energy accelerator experiments).

 Therefore, the modern status of the theory allows us
 to consider the homogeneous approximation, the absolute
 status of measurement standard of the Paris meter,
 the Planck initial data
 $a_I\simeq 10^{-60}$ of origin of the matter, and the
 Higgs field (i.e., material)
  origin of the cosmic evolution of the scale factor as
  subjects of scientific research rather than dogmas.

 In the paper we listed the arguments in favour of that the latest
 observational data fit the relative measurement standard
 (with the expanding  Paris meter) and the vector boson initial data
 $a_I\simeq 10^{-14}$ of origin of the matter, and the gravitational
  origin of the cosmic evolution of the scale factor treated as
  the collective inertial motion along a geodesic in the field space of
  metric components.

  The considered theory explains
  the creation of the universe in the world field space $(a|\bar F)$ with
  the field evolution parameter $a$ and the energy
 $$P_a=E_{universe} + H_{\rm QFT}d\eta/da+..~,$$
 where
 $$E_{universe}=
  2V_0\vh_0^2 (H_Ia_I^2)\log(a_0/a_I)$$
  like the modern quantum field theory explains particle creation in the Minkowski world space $(X_0|X_i)$
  with the energy
  $$P_0=E_{mass} + (P_i^2/2m)(d\eta/dX_0)+..$$
  (where $E_{mass}=mc^2$). In both the cases $d\eta$ means the geometric time
  interval.

  The considered theory introduces the geometric time
  interval in the Hamiltonian description  by the Levi-Civita
  canonical transformation of the field variables to the geometric set of variables.
  In this case, the causal quantization of the scale factor
  removing the negative energy $P_a$ explains both the absence of the
  cosmological singularity $a=0$ (in the field wave function of
  created universe) and
  the positive arrow of the geometric time $\eta$
  (in the geometric wave function of
  created universe) as the consequences of  stability
  of quantum theory.
 The relation of two evolution parameters of these wave functions
 in the form of the evolution of the universe $a(\eta)$
 is  a pure relativistic effect.
  Just this relation gives the dynamic status of the Hubble
 law in the quantum universe.

 The theory explains the origin of observational matter.
 Remind that the collective motion with the momentum
 $P_a$ allows us to define energy of the universe compatible with
 the one in quantum field theory $H_{\rm QFT}$, the observational
 energy density, and observational particles.

 The theory points out the creation of $W,Z$ - vector
  bosons from the geometric vacuum due to their
 mass singularity $\bar m_{W}(a\to 0)\to 0$.
 This mass singularity is the physical origin of
 the temperature of the matter.  The inertial
 cosmic motion in the coset $A(4)/L$ leads to the
 definite temperature of the matter
 $T_I \simeq (m^2_{W}H)^{1/3} K$ depending on
 the boson mass $m_{W}$ and the Hubble parameter $H$.
 This temperature is an integral of the inertial motion
  and this integral coincides
 with the primordial value of the Hubble parameter $T_I=H_I$.

 These primordial bosons decay with the baryon number
 violation~\cite{yaf}. The CMB radiation as
 the product of decay of the primordial bosons keeps
 the primordial value of the temperature
 $T_{CMB } \simeq (m^2_{W0}H_0)^{1/3}=2.7 K$, where $m_{W0},~H_0$ are the
 present-day values of boson mass and the Hubble parameter in the stiff regime.

Thus, we have shown that the quantum universe could be created in
the CMBR reference frame with the value of the primordial scale factor
$a_I \approx 0.3 \times 10^{-14}$ and the primordial Hubble parameter
$H_I \approx 2.7 K$.

\section{Conclusion}

 The treatment of General Relativity as
 the theory of nonlinear representation of the affine group
 helped us to determine the geometry of the field space, and
 extended the principles of relativity
 (including the concepts of the inertial motion, geodesic line, relative
 and absolute "coordinates") to this field space.

 The consistent and complete description of the creation of the quantum
 universe and its evolution allowed us
 to consider the creation of matter in the quantum universe in the
 Standard Model of the electroweak and strong interactions.

This theoretical description depends on the Lorentz frame.
In this sense the explicit relativistic covariance is lost.
Remind that this fact is not a defect of the theory.
The relativistic covariance means that a complete set of states
 obtained by all Lorentz transformations of a state
 in a definite frame of reference
 coincides with a complete set of states
 obtained by all Lorentz transformations of this state
 in another frame of reference
 (see the review of papers by
 V. Bargmann, E.P. Wigner, and A.S. Wightman in the monography~\cite{bww}).

  Moreover, following to Julian Schwinger
 "...we reject all Lorentz
 gauge formulations as unsuited to the role of providing the fundamental operator
 quantization ..."~\cite{sch} counting that any gauges that do not depend
 on a  reference frame are not correct
 for the description of amplitudes of  all processes in quantum theory,
 including the creation of the universe, time, and matter
 with the exception of the narrow class of amplitudes of the scattering of elementary local
 fields on their mass-shell~\cite{f,rf}.

\section*{Acknowledgments}
V.N.P. thanks Prof. W.~Thirring for the discussion of experimental consequences
of General Relativity with relative measurement standard and Profs. B.M.~Barbashov,
C.~Isham and T.W.B. Kibble for discussions of the field nature of evolution parameter
in GR. We are  grateful to , D.~Behnke, D.B.~Blaschke, A.A.~Gusev,
 and S.I.~Vinitsky for fruitful discussions.

 \section*{ Appendix A: Hamiltonian formalism in the coset A(4)/L}

\renewcommand{\theequation}{A.\arabic{equation}}

\setcounter{equation}{0}
  A choice of a Lorentz-frame in GR   means the separation of all
underlined indices into the time-like and space-like ones
$(\underline \mu=\underline 0,\underline a)$.

 To formulate Hamiltonian dynamics in GR,
 besides of the Lorentz-frame in the Minkowski space-time $\underline \mu$
 we should choose also the {\bf Riemannian frame}.
 The Riemannian frame of reference for solving  the evolution problem in GR
 is known as the "kinemetric" frame~\cite{vlad}. This frame
  means the $3+1$ foliation of  the Riemannian space-time
$$
\omega_{\underline 0}=Ndx^0,~~~~~~~~\omega_{\underline a}=e_{\underline a i}
(dx^i+N^idx^0)~,
$$
where $e_{\underline a i}$ is ``drei-bein''.
This foliation in terms of an Einstein interval
\bea \label{dse}
  (ds)^2&=&g_{\mu\nu}dx^\mu dx^\nu= (N dx^0)^2\nonumber\\
&&-
  {}^{(3)}g_{ij}
 \left(dx^i+N^idx^0\right)\left(dx^j+N^jdx^0\right)~~
 \eea
  was applied for the generalized Hamiltonian approach to  the Einstein theory
 of gravitation~(\ref{sGR}) by Dirac
 and Arnowitt, Deser and Misner~\cite{d1,ADM}.
 This foliation keeps all ten components of the metric
 with the lapse function $N(x^0,\vec x)$, three shift
 vectors $N^i(x^0,\vec x)$,
 and six space components ${}^{(3)}g_{ij}(x^0,\vec x)$ depending on the
  coordinate time $x^0$ and  space coordinates $\vec x$.
 The Dirac-ADM parametrization characterizes
 a family of hypersurfaces $x^0=\rm{const.}$ with the unit normal
 vector
 $\nu^{\alpha}=(1/N,-N^k/N)$ to a hypersurface. The second
 (external) form
$$
 \frac{1}{N}\left(\partial_0{}^{(3)} g_{ij}
 - N_{j|i} - N_{i|j}\right)\equiv
$$
\be \label{ext}
 \frac{1}{N}
 \left[\left(\partial_0-N^l\partial_l\right){}^{(3)} g_{ij}
 - {}^{(3)} g_{il}\partial^kN^l-{}^{(3)} g_{jl}\partial_iN^l\right]
\ee
shows how this hypersurface is embedded into the four-dimensional
 space-time. Here $N_{i|j}$ is the covariant derivative
 with respect to the metric ${}^{(3)} g^{kj}$.
In terms of ``drei-bein'' the expression~(\ref{ext}) takes the form
\be \label{ext1}
\left(\partial_0{}^{(3)} g_{ij}
 - N_{j|i} - N_{i|j}\right)\equiv e_{\underline a i}
(D_0e)_{\underline a j}
+(i~\leftrightarrow~j)~,
\ee
 where
\be \label{exto}
(D_0e)_{\underline a i}=\left(\partial_0-N^l\partial_l\right)
e_{\underline a i} -e_{\underline a l}\partial_iN^l~.
\ee
 A gauge group is considered as  the
 group of  diffeomorphisms of the Dirac-ADM parametrization
 of the metric~(\ref{dse})~\cite{vlad}
 \be \label{gt}
 x^0 \rightarrow \tilde x^0=\tilde x^0(x^0);~~~~~
 x_{i} \rightarrow  \tilde x_{i}=\tilde x_{i}(x^0,x_{1},x_{2},x_{3})~,
 \ee
 \be \label{kine}
 \tilde N = N \frac{dx^0}{d\tilde x^0};~~~~\tilde N^k=N^i
 \frac{\partial \tilde x^k }{\partial x_i}\frac{dx^0}{d\tilde x^0} -
 \frac{\partial \tilde x^k }{\partial x_i}
 \frac{\partial x^i}{\partial \tilde x^0}~.
 \ee
 These transformations
 conserve the family of hypersurfaces $x^0=\rm{const.}$, and they
 are called a kinemetric subgroup~\cite{plb,ps1,vlad}
 of the group of general coordinate
 transformations~(\ref{x}).
 The group of kinemetric transformations contains
 reparametrizations of the {\it coordinate time}~(\ref{gt}).
 This means that
 there are no physical instruments that can measure
 this {\it coordinate time}.
 The definition of the kinemetric frame of reference requires
 to point out an {\bf invariant field evolution parameter}
  in the field space $g_{\mu\nu}=(g_{ij},N,N^k)$
 which plays the role of the fourth time-like coordinate in
 Minkowskian space-time in Special Relativity.

The Einstein action in the kinemetric frame takes the form
$$
 S_{\rm GR}[e|\vh_0]=-\int d^4x\sqrt{-\bar g}\frac{\vh_0^2}{6}R(g)=
$$
 \be \label{Econf0}
 \int\limits_{x^0_1 }^{x^0_2 }dx^0\int\limits_{V_0}d^3x
 \left[ \verb"K"(e|\vh_0)-\verb"P"(e|\vh_0)+
 \verb"S"(e|\vh_0)\right]~,
  \ee
 where
 \be \label{Eck0}
 \verb"K"(e|\vh_0)=\frac{\vh_0^2|e|}{24 N} [\pi_{\underline a\underline b}
\pi_{\underline a\underline b}-
  \pi_{\underline b\underline b}\pi_{\underline a\underline a}]
 \ee
 is the kinetic term,
 \be \label{Ecp0}
 \verb"P"(e|\vh_0)=\frac{\vh_0^2
  N |e|}{6}~~{}^{(3)}R(e)
 \ee
 is the potential  term with a three dimensional curvature in terms of ``drei-beins''
 $e_{\underline a i }$, and
 \be \label{EcS0}
 \verb"S"(e|\vh_0)=\frac{\vh_0^2 }{6}\left(\partial_0 - \partial_kN^k\right)
\left( \frac{|e|\pi_{\underline a\underline a}}{N}\right)
 -\frac{\vh_0^2 }{3}
 \partial_i\left(|e| g^{ij}\partial_j N\right)
 \ee
 are the standard
 ADM "surface terms" which do not contribute to the equations of motion.
 Here we used the following definitions
\be\label{def0}
\sqrt{{}^{(3)}\bar g}=det||e_{\underline a i}||~,
~~~
g^{ij}=(e^{-1})_{i\underline a}~(e^{-1})_{j\underline a}~,
\ee
$$
\pi_{\underline a \underline b }=
\omega^R_{\underline a \underline b }(D_0)=\frac{1}{2}
[(D_0e)_{\underline a i}(e^{-1})_{i\underline b}+(
\underline a~\leftrightarrow~\underline b)])~.
$$
The definition of $\omega^R_{(\underline a \underline b )}(D_0)$ is given by
~(\ref{tR}) where $(de)$ is replaced by the covariant derivative~(\ref{exto}).

One can choose a triangle ``drei-bein''
\be \label{triangle}
e_{\underline a i}=0,~~~~~~~~~\underline a< i~
\ee
that is the continuation of the $4=3+1$ foliation for the $3=1+1+1$ one.

 To make the discussion of the {\bf invariant field evolution parameter} in GR
more transparent, it is useful to
separate the determinant of the three-dimensional metric~\cite{Y}
 \be
e_{\underline a i}=\psi^{2}e^T_{\underline a i}~,~~~
det ||e^T_{\underline a i}||=1~,~~
det ||g^T_{ij}||=1.
  \ee
 Then, instead of the Lagrangians (\ref{Eck0}), (\ref{Ecp0}) and (\ref{EcS0})
 we get the kinetic term
 \be \label{Eckm}
 \verb"K"(g|\vh_0)=\frac{\vh_0^2 \psi^6}{ N}
  \left[\frac{\pi^T_{\underline a\underline b}
\pi^T_{\underline a\underline b}}{24}-4\pi^2_{\psi}\right]~,
  \ee
  where
  \be \label{pt}
 \pi_{\psi}=(\partial_0-N^k\partial_k)\log\psi-\frac{1}{6} \partial_kN^k~,
\ee
 the potential term is
 \be \label{Ecpm}
 \verb"P"(g|\vh_0)=\frac{\vh_0^2N \psi^2}{6}\left[{}^{(3)}R(g_{(T)})
 +\frac{8}{\psi}g_{(T)}^{ij}\partial_i\partial_j\psi \right],
 \ee
 and  the "surface" term is
 \be \label{EcSm}
 \verb"S"(g|\vh_0)= {\vh_0^2 }{2}\left(\partial_0 - \partial_kN^k\right)
\left( \frac{\psi^6\pi_{\psi}}{N}\right)
\ee
$$
-\frac{\vh_0^2 }{3}g^{ij}_{(T)}
 \partial_i\left(\psi^2\partial_j N\right)~.
$$
Now we can introduce canonical momentum
using the standard definitions and eqs.~(\ref{def0}),~(\ref{Eckm})
\be \label{momp}
P_{\psi}=\frac{\partial
{\rm K}
(g|\vh_0)}{\partial (\partial_0\psi)}
=-8\vh_0^2 \psi^5\pi_{\psi}~,
\ee
\be \label{mome}
P_{\underline a\underline b}=\frac{1}{2}\left[e^T_{\underline a i}
P^i_{\underline b}+({\underline a}\leftrightarrow {\underline b})
\right]~,
\ee
$$
P^i_{\underline a}=
\frac{\partial{\rm K}(g|\vh_0) }{\partial (\partial_0e_{\underline a i})}
=\frac{\vh_0^2 \psi^6}{12 N}
  \left[(e^{-1}_T)_{\underline a i}\pi^T_{\underline a\underline b}\right]~.
$$
In terms of these momenta the Einstein
action  takes  the first-order Hamiltonian form
\be\label{hamc}
S_{\rm GR}[F|\vh_0]=S_F+S_{\rm GR}=
\ee
$$
\int dx^0 d^3x
\left[\sum\limits_{F=\psi,e^T}P_F\partial_0F
-N{\cal H}+N^k {\cal P}_k +\verb"S"(g|\vh_0)\right],
$$
where the local Hamiltonian density
\be\label{t00}
N{\cal H}=\verb"K"(g|\vh_0)+\verb"P"(g|\vh_0)
\ee
is the sum of terms defined
by eqs.~(\ref{Eckm}),~(\ref{Ecpm}), the kinetic Lagrangian
\be
\verb"K"(g|\vh_0)=N\left\{\frac{6}{\vh^2_0\psi^6}[
 P_{(\underline a\underline b)}  P_{(\underline a\underline b)}]-
\frac{1}{16\vh^2_0\psi^4}P_{\psi}^2\right\}
\ee
 depends on the canonical momenta $P_{(\underline a\underline b)}$
and $P_{\psi}$,  and
\be\label{t0kp}
 {\cal P}_k=\partial_j(P^i_{\underline a }e^T_{\underline a k })+
 P^i_{\underline a }\partial_ke^T_{\underline a i }
-T^0_k
\ee
is the local momentum where the energy momentum tensor
that depends on the trace of the second external form
 \be \label{t0k}
T^0_k=
P_{\psi}\partial_k \psi-\frac{1}{6}\partial_k (P_{\psi}\psi)
\ee
is distinguished.

We have the Dirac generalized Hamiltonian system with
four local constraints~(\ref{t00}),~(\ref{t0kp})
\be \label{fc}
{\cal H}_{\rm tot}=0, ~~~~~~~~~~~~{\cal P}_{k}=0~.
\ee
In terms of ``observables''
$$
{\cal P}_{\underline a  }={\cal P}_{k}(e^{-1}_T)_{ k \underline a }
$$
the second three constraints ${\cal P}_{\underline a  }=0$
take the transparent form of the transverseness condition
\be \label{tc}
\partial_j P^j_{\underline b  }=\tilde T^0_{\underline b  }~,
\ee
where
$$
\tilde T^0_{\underline b  }=T^0_k(e^{-1}_T)_{ k \underline b }+
 P^j_{\underline a  }(\partial_ke^T_{ \underline a j }-
\partial_je^T_{ \underline a k })(e^{-1}_T)_{ k \underline b }~.
$$

Four constraints~(\ref{fc}) should be accompanied by four gauges.
Dirac~\cite{d1} chose the so-called  minimal embedding of a three-dimensional
hypersurface into the four-dimensional space-time
\be\label{mind}
P_{\psi}=0
\ee
and the transverseness condition that in terms of "drei-beins"
takes the form
\be\label{trd}
f_{\underline b}(e^T)=0~\Rightarrow~\partial^ie^T_{ \underline b i}=0
\ee
compatible with the constraint~(\ref{tc}).
The minimal embedding allowed Dirac to remove all local excitations of
metric with the negative norm and a negative contribution to energy.

Faddeev and Popov~\cite{fp2}  used  other gauges. Using
general coordinate transformations they
removed  all components of the three-dimensional metric besides two.
Using general coordinate transformations one can also remove all components
of "drei-beins" besides  two.
In particular,  we can keep two nondiagonal components
in the triangle basis~(\ref{triangle}) of $(3=1+1+1)$ foliation
\be\label{tria1}
e^T_{\underline a i}=
\left| \begin{array}{llcl}
1 & 0 & 0 \\
e_{\underline 2 1} & 1 & 0 \\
0 & e_{\underline 3 2} & 1
\end{array}\right|~.
\ee
The triangle basis~(\ref{tria1}) allows us to find the polynomial form of
the three-dimensional metric
\be\label{tria2}
g^T_{ij}= \left|
\begin{array}{llcl}
1 & e_{\underline 2 1} & 0 \\
e_{\underline 2 1} & 1+e_{\underline 2 1}^2 & e_{\underline 3 2} \\
0 & e_{\underline 3 2} & 1+e_{\underline 3 2}^2
\end{array}\right|~,
\ee
\be\label{tria3}
g^{ij}_T=
\left| \begin{array}{llcl}
1+e_{\underline 2 1}^2(1+e_{\underline 3 2}^2) & -e_{\underline 2 1}(1+e_{\underline 3 2}^2) & 0 \\
-e_{\underline 2 1}(1+e_{\underline 3 2}^2) & 1+e_{\underline 3 2}^2 & -e_{\underline 3 2} \\
e_{\underline 2 1}e_{\underline 3 2} & -e_{\underline 3 2} & 1
\end{array}\right|
\ee
for which the GR action becomes polynomial.
In these cases, with the affine version of General Relativity one hopes
to prove its renormalizability.

Thus, the Hamiltonian action~(\ref{hamc}) with all constraints takes the form
\be\label{hall}
S_{\rm GR}[F|\vh_0]=
\ee
$$
\int dx^0\left\{
\left[ \int d^3x \sum\limits_{F=\psi,e,}P_F\partial_0F\right]-
H_{\rm tot}[F|\vh_0]\right\}~,
$$
where
\be\label{hall1}
H_{\rm tot}[F|\vh_0]=
\ee
$$\int d^3x\left[
N{\cal H} -N^k {\cal P}_k + C_0P_{\psi}+C_{\underline b}f_{\underline b}(e^T)-
\verb"S"\right]
$$
is the total Hamiltonian with
the Lagrangian multipliers $N, N^k, C_0, C_{\underline b}$.

In the context of our consideration of the problem of energy and  time
it is very important to emphasize the positive contribution  of the
transverse "drei-beins" to the Hamiltonian density ${\cal H}_{\rm tot}=0$
in the conventional perturbative theory.  The latter
begins with the gauge
\be\label{gauge2}
N=1,~~~~~~~~~~~\psi=1
\ee
that supposes that the measurable time is identified with
the coordinate time $x^0$.
We see below that the gauge $N=1$
 removes part of pure  relativistic results including definition of
nonzero energy of the proper time, a positive arrow  of the
 time, and cosmic initial data.

 The perturbation theory with $\psi=1$ loses all cosmology.
Recall that the evolution of the universe  is identified with the
determinant of the spatial metrics in the homogeneous approximation
$\psi^2=a(x^0)$ that is well known as cosmic scale
factor.

\end{document}